\documentclass[%
nofootinbib, 
 reprint,
 preprintnumbers,
 amsmath,amssymb,
 aps,
 pra,
]{revtex4-1}

\usepackage{mathtools}
\usepackage{graphicx}
\usepackage{dcolumn}
\usepackage{bm}
\usepackage{braket}
\usepackage{bm}

\usepackage{mathrsfs}
\usepackage{bbm}

\usepackage{xcolor}
\usepackage{comment}
\usepackage{lipsum}

\begin{document}

\preprint{KEK-TH-2123}

\title{On the Validity of Weak Measurement Applied for Precision Measurement
}

\author{Yuichiro Mori}
\email{yuichiro@post.kek.jp}
\affiliation{%
 Department of Physics, The University of Tokyo, Tokyo 113-0033, Japan.
}%
\author{Jaeha Lee}
\email{lee@iis.u-tokyo.ac.jp}
\affiliation{%
 Institute of Industrial Science, The University of Tokyo, Chiba 277-8574, Japan.
}%
\author{Izumi Tsutsui}
\email{izumi.tsutsui@kek.jp}
\affiliation{%
 Theory Center, Institute of Particle and Nuclear Studies, High Energy Accelerator Research Organization (KEK), Ibaraki 305-0801, Japan.
}%

\date{\today}

\begin{abstract}
{
We present a general framework of examining the validity of weak measurement -- the standard procedure to acquire Aharonov's weak value -- which has 
been used intensively in recent years for precision measurement, taking advantage of the amplification mechanism available for the weak value.  Our framework allows one to deal systematically with various causes of uncertainties intrinsic to the actual measurement process as well as those found in the theoretical analysis employed to describe the system.  Using our framework, we examine in detail the two seminal experiments,  Hosten's detection measurement of the spin Hall effect of light \cite{Hosten_2008} and Dixon's ultra sensitive beam deflection measurement \cite{Dixon_2009}.  Our analysis shows that their results are well within the range of amplification (actually in the vicinity of the optimal point) where the weak measurements are valid.  This suggests that our framework is both practical and sound, and may be useful to determine beforehand the possible extent of amplification in the future weak measurement experiments.
}
\end{abstract}

\pacs{Valid PACS appear here}

\maketitle

\section{Introduction}\label{sec:introduction}

In pursuing a time-symmetric description of measurement process, Aharonov {\it et al.}~proposed in 1964 a formalism 
in which one specifies the process not only by the initial state $\ket{\phi}$ but also by the final state $\ket{\psi}$, thereby rendering the whole description free from a particular direction of time \cite{Aharonov_1964}.   If a physical quantity $A$ is observed in the process \lq weakly\rq\ so that the final state is kept intact, the value obtained is expected to be characteristic to the process.   
This may sound a bit odd but it is perfectly consistent with the standard quantum mechanics, since all it amounts to is simply  
to look at the amended process of measurement in which we measure $A$ for the initial state $\ket{\phi}$ as usual but we keep the obtained data only when the state is confirmed to be $\ket{\psi}$ at the end of the process.  Because of our freedom in choosing both the initial and final states (which is not necessarily the time-evolved state from the initial state) in advance, the two states, $\ket{\phi}$ and $\ket{\psi}$, are called {\it preselected state} and {\it postselected state}, respectively. 

Apart from providing a time-symmetric viewpoint in quantum mechanics, this description acquired a notable significance later in 1988 when Aharonov, Albert and Vaidman \cite{Aharonov_1988} arrived at a novel physical quantity called {\it weak value},
\begin{equation}\label{def:weakvalue}
A_{\rm w} := \frac{\langle \psi | A | \phi \rangle}{\langle \psi | \phi \rangle}.
\end{equation}
This is the value of the physical quantity $A$ obtained in the weak limit of the measurement interaction, dubbed {\it weak measurement}, and is characteristic to the process specified in the time-symmetric description.
The significance of the \lq discovery\rq\ of the weak value is twofold.  First, it furnishes a useful tool in considering the foundations of quantum mechanics, shedding a new light on 
quantum paradoxes many of which are concerned with the reality (ontological aspect) of the observable $A$.  
These include trajectories in the double slit experiment \cite{Kocsis1170, Mori2015}, 
the three box paradox \cite{Aharonov_1991}, Hardy's paradox \cite{Aharonov_2002} and the Cheshire cat effect \cite{Aharonov_2013}, all of which have been demonstrated experimentally as well \cite{Yokota_2009, Denkmayr:2014aa, Okamoto:2016aa}.  More recently, the weak value allows one to take backward evolving states into consideration in an attempt to resolve a paradox involving the existence of photons in certain paths in a nested interferometer \cite{Danan_2013}.  Active discussions are still ongoing in these respects (see, for instance, \cite{Vaidman_2018} and the references therein), while new proposals on the use of the weak value are also being made for different issues such as neoclassical realism \cite{1808.09951}, nonlocality \cite{1810.05039} and complementarity \cite{Thekkadath_2018}.

Secondly, and perhaps more importantly in practical viewpoints, the weak value offers a novel method of precision measurement by means of its ability of amplification.  
Indeed, on account of the freedom in choosing the two (preselected and postselected) states $\ket{\phi}$ and $\ket{\psi}$, we may
amplify the weak value $A_{\rm w}$ in (\ref{def:weakvalue}) at will -- even exceeding the range of spectrum of the observable $A$ -- by rendering the denominator $\langle \psi | \phi \rangle$ (vanishingly) small while maintaining the numerator finite.   This was originally pointed out in \cite{Aharonov_1988} for the case of spin $A = \sigma_x$ for which the value $(\sigma_x)_{\rm w} = 100$ was argued to be realized, but the first decisive demonstration of amplification appeared two decades later when Hosten and Kwiat showed that the spin Hall effect of light (SHEL) which is extremely tiny can be detected via the weak value amplification \cite{Hosten_2008}.  This was soon followed by another skillful experiment by Dixon {\it et al.}~which detected ultra sensitive beam deflection (USBD) using the same technique \cite{Dixon_2009}.  These two experimental demonstrations on the use of weak value amplification for precision measurement stood as a landmark and paved the way to its application to precision measurement of a variety of other physical properties, such as phase, frequency or temperature (for a review, see {\it e.g.}~\cite{Dressel_2014}).  In recent years, the range of application has broadened considerably to the extent that one deals with massive particles like trapped ions \cite{1811.06170} and may even envisage observation of gravitational waves \cite{NAKAMURA_2018} or measurement of gravitational constant with unprecedented accuracy \cite{Kawana_2018}.  

These, however, do not come for free.  In fact, the price we pay for amplifying the weak value is that the rate of passing the screening at postselection  (given by $\vert \langle \psi | \phi \rangle\vert^2$ in the weak limit) is diminished, which implies that we obtain only scanty data if we amplify too much, causing unwanted noise to invalidate what we have measured.  More precisely, one can show under certain assumptions that the signal to noise ratio of the weak measurement for detecting the weak coupling (the strength of the weak interaction) remains constant \cite{Knee_2014}, and that the scale of amplification is limited due to the nonlinear effect in the weak measurement \cite{koiketanaka_11}.   However, all these analyses are based on some models of noise and/or disturbance despite that these elements are intrinsically difficult to pin down in practice.  This is a serious problem, because it indicates that, strictly speaking, we do not know for sure if the results of our weak measurements, including those already performed which employ the technique of weak value amplification, are reliable or not.  We have, therefore, led to the situation where some, definite and desirably general, point of reference is needed to examine the issue.  In other words, given a set of parameters characterizing the measurement and the theoretical procedure to be used for analysis, we wish to know how much we can amplify the weak value without destroying the validity of measurement.   

Motivated by this demand, in our earlier work \cite{Lee_2014} we proposed a general framework to deal with measurement uncertainty in the weak measurement in which the trade-off relation between the amplification factor and the level of uncertainty can be discussed explicitly.  In the present paper, we extend it so that the nature of the uncertainties becomes more transparent when it is decomposed into elements of different theoretical/experimental origins encountered in the actual measurement process.  Armed with this, we analyze the validity of the two aforementioned seminal  experiments, the SHEL measurement \cite{Hosten_2008} and the USBD measurement \cite{Dixon_2009} in detail.
The outcome is assuring, that is, these two measurements are indeed valid and hence their measured values are reliable.  Curiously enough, we also find that the amplification scale used in their experiments are almost optimal in the light of our analyses in which the significance of the observed results is examined with respect to the possible uncertainties in each of the experiments.  In the absence of a definite framework which specifies completely the uncertainties which are evasive in nature, we believe that, at least, our results suggest the soundness and consistency of our framework along with the validity of the two experiments examined with it. 

The present paper is organized as follows.  We first provide in Sect.2 a concise review on both the standard von Neumann measurement and the weak measurement.  This is to make our paper self-contained so that anyone who is interested in the possible application of weak measurement but unfamiliar with the subject can go through our subsequent arguments without feeling much difficulty.  We then present in Sect.3 our basic theoretical framework of analysis on the weak measurement in which a classification of errors/uncertainties and our criterion on the validity of measurement are given.  In Sect.4, the two experiments are then put into our framework, one by one, to see if each of them is actually meaningful as a precision measurement.  Sect.5 is devoted to our conclusion and discussions.   Several appendices are included to supplement our arguments in the main text.

\section{Theoretical Framework}\label{sec:theoretical_framework}

\subsection{Standard (von Neumann) measurement}

We start by recapping the standard indirect measurement briefly.  
Let $\ket{\psi}$ be the state of the system for which we want to measure the physical observable $A$.
All the states of the system form a Hilbert space $\mathcal{H}$ and the observable $A$ is a self-adjoint operator acting
in $\mathcal{H}$.  The aim of the measurement is to obtain some sort of information about the system and/or the observable
when they are not precisely known, as one typically encounters in the search of novel phenomena for which no definite Hamiltonian is given.
The most fundamental object of our concern for this is to find the expectation value $\bra{\psi}\!A\! \ket{\psi}$ of the observable $A$ from which
the desired information may be retrieved.

The standard quantum measurement, often referred to as the von Neumann measurement, is an indirect measurement procedure in which we consider 
an auxiliary meter system $\mathcal{K}$ in addition to the target system $\mathcal{H}$.  
For the meter system $\mathcal{K}$, it is enough for our purposes to take the simple one dimensional quantum mechanical system
given by the space of square-integrable functions $\mathcal{K}=L^{2}(\mathbb{R})$, where a pair of observables $\hat x$ and $\hat p$ satisfying the canonical commutation relation $[\hat x, \hat p] = i\hbar$ are equipped.

Now, for the measurement we need to transfer the information of the observable contained in the target system to the meter system by creating a proper entanglement between the two systems.  The interaction of the von Neumann measurement to fulfill this mission is furnished by the unitary operator acting in the total system $\mathcal{H} \otimes \mathcal{K}$,
\begin{equation}\label{def:unitary_interaction}
U(\theta):= e^{-i \frac{\theta}{\hbar} A \otimes Y},
\end{equation}
where $\theta$ is a real parameter representing the coupling of the target system and the meter system, and
$Y$ is an observable operator of the meter system chosen to be either $Y=\hat x$ or $Y= \hat p$.  For brevity,  throughout this paper we omit the symbol of hat on operators denoted by capital letters such as $A$ and $Y$.  

To see that this scheme achieves our mission, let $\ket{\chi}$ be the meter state prepared initially along with the system state $\ket{\psi}$.
Then, under the measurement interaction \eqref{def:unitary_interaction}, the state $\ket{\Psi} := \ket{\psi} \!\ket{\chi}$ of the total system undergoes the change,
\begin{equation}
\label{def:comopsite_state}
\ket{\Psi} \to \ket{\Phi} := U(\theta) \ket{\psi}\! \ket{\chi}.
\end{equation}
We may choose
an observable $X$ of the meter system satisfying $[X, Y] = i\hbar$ and regard it as the position of the meter.
We then observe that during the change \eqref{def:comopsite_state}
the expected position shifts by (see Appendix \ref{Conventional Measurement})
\begin{align}
\label{shiftone}
\Delta(\theta) 
&:= \braket{\Phi| I\otimes X |\Phi} - \braket{\Psi| I\otimes X |\Psi}  \nonumber \\
&\phantom{:}= \theta\braket{\psi|A|\psi},
\end{align}
where $I$ is the identity operator in the target system $\mathcal{H}$.   We now can retrieve the expectation value of the observable $A$ directly from the shift of the meter as $\Delta(\theta)/\theta = \braket{\psi|A|\psi}$.

\subsection{Weak Measurement}\label{sec:conditioned_measurement}

The basic idea of weak measurement is that, in addition to the conventional process called {\it preselection} in which we prepare the initial state $\ket{\psi}$ for the system, 
we also
implement the process called {\it postselection} before we observe the meter system in the weak coupling regime where $\theta$ is small.  

The postselection
is the conditioning of the state of the system after the interaction.  Given a state $\ket{\phi}$ of the system chosen for the postselection, we examine if the state of the system is $\ket{\phi}$ or not after the measurement interaction \eqref{def:unitary_interaction}.  
Observation of the meter system $\mathcal{K}$ is carried out only when the result of the examination is affirmative, in which case the (normalized) state of the meter system becomes
\begin{align}
\label{postselst}
\ket{\xi} = \frac{\bra{\phi}U(\theta)\ket{\psi} |\chi\rangle}{||\bra{\phi}U(\theta)\ket{\psi}\ket{\chi}||}.
\end{align}
If we further allow a possible time development of the meter system described by the unitary operator $V_1$ before its observation, we may write the final state of the meter as
\begin{align}
\label{zeta}
\ket{\zeta} = V_1\ket{\xi}.
\end{align}

In this procedure, the shift of the expectation value of the meter observable $X$, now chosen arbitrarily, 
\begin{align}
\label{def:shift}
\Delta_{\rm w}(\theta) := \braket{\zeta|X|\zeta} -  \braket{\chi |X|\chi}  
\end{align}
is given by 
the formula (for the derivation, see Appendix \ref{Weak Measurement}),
\begin{align}
\label{def:shift3}
\Delta_{\rm w}(\theta) &= \braket{\chi|V^{\dag}_1 X V_1|\chi} -  \braket{\chi |X|\chi}  \nonumber \\
&\qquad + \theta \cdot \frac{2}{\hbar}{\rm Im} \left[ A_{\rm w}  {\rm C}_t \right] + O(\theta^2),
\end{align}
where
$A_{\rm w}$ is the weak value given in (\ref{def:weakvalue}), 
and we have defined a {\it modified covariance} by
\begin{equation}\label{def:covariance}
{\rm C}_t := \braket{\chi |\left(V^{\dag}_1 X V_1 - \braket{\chi|V^{\dag}_1 X V_1|\chi}\right) \left(Y - \braket{\chi| Y |\chi}\right)|\chi}.
\end{equation}

Now, for simplicity we restrict ourselves to the case where the time development described by $V_1$ does not alter the expectation
value of the meter observable $X$, {\it i.e.}, 
\begin{equation}\label{chicond}
\braket{\chi|V^{\dag}_1 X V_1|\chi} =  \braket{\chi |X|\chi},
\end{equation}
under the state $\ket{\chi}$ prepared for the meter.
We then find that in the linear approximation the shift becomes
\begin{align}
\label{def:shift2}
\Delta_{\rm w}^{(1)}(\theta) &:= \theta \cdot \frac{2}{\hbar}{\rm Im} \left[ A_{\rm w}  {\rm C}_t \right] \nonumber \\
&\phantom{:}= \theta \cdot \frac{1}{\hbar}\Big\{{\rm Re} \left[ A_{\rm w}\right]  \frac{1}{i}\braket{\chi |[V^{\dag}_1 X V_1, Y] |\chi} \nonumber \\
&\!\!\!\!\!\!\!\!\!\!\!\!\!\!\!\!\!\! + {\rm Im} \left[ A_{\rm w}\right]  \big(\braket{\chi |\{V^{\dag}_1 X V_1, Y\} |\chi} - 2\braket{\chi |X|\chi}\braket{\chi |Y|\chi}\big) \Big\}.
\end{align}

In the actual measurement, as in the standard measurement, 
we usually consider the meter system $\mathcal{K}=L^{2}(\mathbb{R})$ with the 
observable $X = \hat x$ representing the position of the pointer of the meter.  We may also have, for some experiments including those
we analyze later, the time development described by the effective unitary operator of the form,
\begin{equation}\label{vform}
V_1 = e^{\frac{i}{\hbar}\frac{\hat p^2}{2m} t},
\end{equation}
with real constants $m$ and $t$ which may not represent the mass and time associated with the state, yielding 
\begin{align}
\label{Heisenberg-position}
V^{\dag}_1 X V_1 = \hat x - \frac{1}{m}\, \hat p t.
\end{align}  
In addition, it is customary to prepare the Gaussian beam for the measurement so that the centeral position of the beam acts as the pointer of the meter.
One possible case of this is that, at time $-t_0 < 0$, we provide the meter state $\ket{\chi_0}$ prior to the measurement by $\braket{x|\chi_0} = G(x)$ where
$G(x)$ is the Gaussian profile function,
\begin{equation}\label{def:Gaussian_function}
G(x) := \frac{1}{(2\pi a^{2})^{\frac{1}{4}}} e^{-\frac{x^{2}}{4a^{2}}},
\end{equation}
which evolves into the state $\ket{\chi}$ at the time of the measurement $t = 0$, that is, one may write
\begin{align}
\label{prechi}
\ket{\chi} = V_0\ket{\chi_0},
\end{align}
with some unitary operator $V_0$.  Another case may be that the Gaussian profile \eqref{def:Gaussian_function} is realized at $t = 0$ by the state $\ket{\chi}$ with properly chosen $\ket{\chi_0}$ and $V_0$.  In any case, 
when the condition \eqref{chicond} is fulfilled for $\ket{\chi}$, we still maintain the formula of shift \eqref{def:shift2}.
The form of the unitary operator $V_0$ varies, analogously to the operator $V_1$, depending on the measurement we perform,
and it may well be the case that the direct computation of the shift $\Delta_{\rm w}(\theta)$ is done most efficiently by using \eqref{def:shift3} with those unitary factors.

The salient feature of the weak measurement is that the shift $\Delta_{\rm w}(\theta)$ in the linear order can, in principle, be made as large as we want by
choosing appropriately the combination of the preselection and the postselection.  This is so because, as seen in \eqref{def:shift2}, the linear shift $\Delta_{\rm w}(\theta)$ is proportional to the weak value
$A_{\rm w}$ which can be made large by rendering its denominator small while keeping the numerator finite in \eqref{def:weakvalue}.  
A more detailed argument as to how this can be achieved will be found in Appendix~\ref{app:quantum_conditional_expectation}. 
This is the reason why the weak measurement may be useful for precision measurement, as mentioned in the Introduction\footnote{ 
However, we also point out that the na{\"i}ve speculation that the shift $\Delta_{\rm w}(\theta)$ can be made arbitrarily large is generally not true in the full order, since the higher-order terms of the expansion cannot be ignored due to the fact that they may hamper the enlargement for finite $\theta$ \cite{Lee_2016_PTEP}.}.

\section{Measurement Uncertainty}\label{sec:measurement_uncertainty}

To analyze the validity of experiments, and also to find the 
characteristics of the weak measurement, we need to discuss the measurement uncertainty with respect to the amplification available there. 
In the measurement, suppose that we acquire $n$ outputs from the detector by measuring the observable $X$ under the meter state $\ket{\zeta}$ in \eqref{zeta} obtained after the postselection, and let $\tilde{x}_{i}$, $i = 1, \ldots, n$, be the values of the outputs.  
We then have the experimental average,
\begin{align}
\label{expav}
\tilde{X}_{n}=\frac{1}{n}\sum_{i=1}^{n}\tilde{x}_{i},
\end{align}
which is to be compared to the expectation value $\braket{\zeta|X|\zeta}_{\rm app}$ evaluated by using some approximation method in a given 
theoretical model of the total system.   

\begin{figure}
\centering
\includegraphics[width=8.5cm]{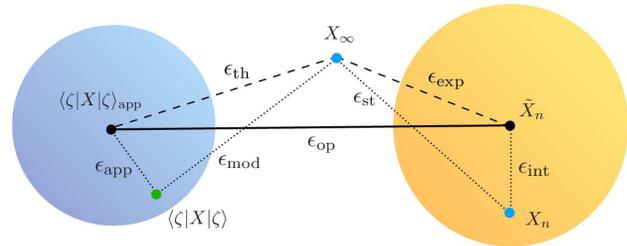}
\caption{The experimental result $\tilde{X}_{n}$, as well as the theoretical value $\braket{\zeta|X|\zeta}_{\rm app}$ estimated with some approximation, may differ from the ideal value $X_\infty$ which would be obtained had we gained infinitely many outputs with perfect accuracy.  
The \lq distance\rq\ between $\tilde{X}_{n}$ and $\braket{\zeta|X|\zeta}_{\rm app}$, which forms the operational difference $\epsilon_{\mathrm{op}}$, is bounded from above by the sum of the theoretical error $\epsilon_{\rm th}$ and the experimental error $\epsilon_{\mathrm exp}$.  Further, 
the theoretical error $\epsilon_{\rm th}$ is decomposed into the approximation error $\epsilon_{\rm app}$ and the model error $\epsilon_{\rm mod}$, while the experimental error $\epsilon_{\rm exp}$ is decomposed into the intractable error $\epsilon_{\rm int}$ and 
the statistical error $\epsilon_{\rm st}$.   Among them, $\epsilon_{\rm app}$ and $\epsilon_{\rm int}$ reside within their own ranges of uncertainties determined from the intrinsic property of our theoretical analysis and our procedure of measurement which are illustrated by \lq balls\rq\ with $\braket{\zeta|X|\zeta}_{\rm app}$ and $\tilde{X}_{n}$ in their respective centers.
}\label{fig:unc_image}
\end{figure}

\subsection{Experimental and Theoretical Errors}

Now, let us consider the difference between the experimentally obtained average value and the theoretically obtained approximated expectation value,
\begin{align}
\label{deftotalunc}
\epsilon_{\rm op}(n) := \left|\tilde{X}_{n}-\braket{\zeta|X|\zeta}_{\mathrm{app}}\right|.
\end{align}
The point is that, besides the fact that this is obtained directly from our experiment and the theory we use and hence it is an operationally meaningful quantity, it can be decomposed into a number of errors of distinctive natures intrinsic to the measurement and the theoretical procedure we follow.

To see this, let us consider the \lq true\rq\ value ${x}_{i}$, $i = 1, \ldots, n$, which we would obtain when the experiment was carried out ideally, that is, with perfect accuracy.  These values give us the average,
\begin{align}
{X}_{n}=\frac{1}{n}\sum_{i=1}^{n}{x}_{i},
\end{align}
which, when the experiment is repeated infinite times, tends to the limit,
\begin{align}\label{infx}
X_\infty = \lim_{n \to \infty} {X}_{n}.
\end{align}
This limiting value is expected to coincide with the theoretical expectation value $\braket{\zeta|X|\zeta}$ 
when a perfectly accurate theoretical model is available.

We then classify the difference \eqref{deftotalunc} in two components of errors according to the inequality,
\begin{align}
\epsilon_{\rm op}(n) 
&=\left|\tilde{X}_{n} - {X}_{\infty} + {X}_{\infty} -\braket{\zeta|X|\zeta}_{\mathrm{app}}\right|
\nonumber\\
&\leq \epsilon_{\rm exp}(n) + \epsilon_{\rm th},
\end{align}
where we have defined the {\it experimental error},
\begin{align}\label{expun}
\epsilon_{\rm exp}(n) := |\tilde{X}_{n}-X_\infty|,
\end{align}
whose meaning is obvious, 
and the {\it theoretical error},
\begin{align}\label{theoun}
\epsilon_{\rm th} := \left|\braket{\zeta|X|\zeta}_{\rm app}-X_{\infty}\right|,
\end{align}
which is understood on the grounds that we have $X_{\infty} = \braket{\zeta|X|\zeta}$ when the theoretical model is perfect and that $\braket{\zeta|X|\zeta} = \braket{\zeta|X|\zeta}_{\rm app}$ when the approximation is accurate or unnecessary.
Each of the two errors is further analyzed as follows.

First, the experimental error may be decomposed into two distinct components according to the inequality,
\begin{align}
\epsilon_{\rm exp}(n) \leq \epsilon_{\rm int}(n) + \epsilon_{\rm st}(n),
\label{expuncertainty}
\end{align}
with the {\it intractable error},
\begin{align}
\epsilon_{\rm int}(n) := |\tilde{X}_{n}-X_{n}|,
\label{intuncertainty}
\end{align}
which originates from the so-called systematic error of the measurement device but also from unknown sources inherent to the environment, 
and the {\it statistical error},
\begin{align}
\epsilon_{\rm st}(n) := |X_{n} - X_\infty|.
\label{stuncertainty}
\end{align}
These two components are to be dealt with in more detail when the experimental setups are specified explicitly.  

Second, the theoretical error \eqref{theoun} can also be decomposed into two distinct components,
\begin{align}
\epsilon_{\rm th} \leq  \epsilon_{\rm mod} + \epsilon_{\rm app},
\label{thuncertainty}
\end{align}
with 
the {\it model error},
\begin{align}
\label{modun}
\epsilon_{\rm mod} := \left|\braket{\zeta|X|\zeta}-X_{\infty}\right|,
\end{align}
and the {\it approximation error},
\begin{align}
\label{appun}
\epsilon_{\rm app} := \left|\braket{\zeta|X|\zeta}_{\rm app}-\braket{\zeta|X|\zeta}\right|.
\end{align}
When the theoretical model of description over the measurement interaction, time development, and the states used for both the preselection and postselection, is exact, the model error vanishes due to the law of large numbers.  

To sum up, these four types of errors provide the upper bound for the operational difference (see Fig.\ref{fig:unc_image}),
\begin{align}
\label{totalunc}
\epsilon_{\rm op}(n) \leq \epsilon_{\rm int}(n) + \epsilon_{\rm st}(n) + \epsilon_{\rm mod} + \epsilon_{\rm app}.
\end{align}

\subsection{Errors under the Fluctuation of the Output Number}

At this point, we need to take account of the actual experimental situation of weak measurement where we cannot specify the number $n$ of 
outputs on account of the fact that the positive results of postselection are not guaranteed
for which we retrieve the meter variable $X$.  One may cope with this situation by resorting to the procedure where one performs 
a fixed number $N$ of trials of postselection and considers
the probability distribution $P(n)$ of the number $n$ of getting positive results out of $N$, where  $\sum_{n = 0}^N P(n) = 1$.
This yields the average of the measured meter position, 
\begin{align}
\tilde X_{N} := \sum_{n = 1}^N P(n)\, \tilde X_{n}.
\end{align}
With this, we may define the operational difference averaged over the fluctuation, 
\begin{align}
\label{deftotalunc2}
\bar \epsilon_{\rm op} := \left|\tilde{X}_{N}-\braket{\zeta|X|\zeta}_{\mathrm{app}}\right|.
\end{align}
Obviously, if we repeat our same argument given above, we end up with the inequality,
\begin{align}
\label{totalunc2}
\bar \epsilon_{\rm op} 
\leq \bar \epsilon_{\rm int} 
+ \bar\epsilon_{\rm st} + \epsilon_{\rm mod} + \epsilon_{\rm app},
\end{align}
where now we have
\begin{align}
\bar \epsilon_{\rm int} := |\tilde{X}_{N}-X_{N}|,
\label{intuncertainty2}
\end{align}
with 
\begin{align}
X_{N} := \sum_{n = 1}^N P(n)\, X_{n},
\end{align}
and
\begin{align}
\bar \epsilon_{\rm st} := |X_{N} - X_\infty|.
\label{stuncertainty2}
\end{align}

In the present case of weak measurement, the probability distribution $P(n)$ is furnished in the following manner.
Let $q$ be the success rate of getting positive output in the postselection,
\begin{equation}
\label{postselcrate}
q := ||\bra{\phi}U(\theta)\ket{\psi}\ket{\chi}||^2.  
\end{equation}
Then, the probability distribution of the success number $n$ is described by the binomial distribution,
\begin{equation}
\label{def:binom_distr}
P(n) = \mathrm{Bi}(n; N,q) :=  \binom{N}{n}\, q^{n} (1-q)^{N-n},
\end{equation}
where 
\begin{equation}
\label{def:binom_coeff}
\binom{N}{n} := \frac{N!}{n!(N-n)!},
\end{equation}
with $n = 0, \ldots, N$.

\subsection{Uncertainty Analysis}

In order to evaluate the four types of errors introduced above, 
each of the errors needs to have an upper bound, which we call {\it uncertainty}, pertinent to the actual settings
of the experiment and the theoretical procedures we employ.   We here provide those uncertainties with some generality to the extent that they suit the two, USBD and SHEL, experiments we analyze later.

\subsubsection{Intractable Uncertainty}

The intractable error mentioned above derives from the systematic error, finiteness of resolution and possibly any other factors that may arise due to the environmental fluctuations.   In the present discussion, however, we assume for simplicity that after careful calibration the systematic error has been eliminated or can be ignored safely when compared to other errors.  In fact, in the two experiments which we analyze later, the systematic uncertainty appears to be suppressed considerably well so that no particular attention to it is necessary in the final analysis.   

We are thus left with the error that may arise from other causes for which we have virtually no control.  To take account of this intractable factor, we 
employ a simple approach, in which we consider a uniform level of doubt associated with each output $\tilde{x}_{i}$ of measurement due to the possible imperfection.  This situation may be described by introducing a real number $\delta_{\rm int} \geq 0$ representing the degree of uncertainty around the measured outcome $\tilde x_{i}$, with the idea that  the true value should lie somewhere in the interval,
\begin{equation}
x_{i} \in \tilde{x}_{i} + [-\delta_{\rm int},\, \delta_{\rm int}].\label{eq:intractable_uncertainty}
\end{equation}
In this model, the uncertainty of the average of the outputs fulfills the relation,
\begin{align}
\epsilon_{\rm int}(n) = \left|\tilde{X}_{n}-X_{n}\right|=\left|\sum_{i=1}^{n}\frac{\tilde{x}_{i}-x_{i}}{n}\right| \leq\delta_{\rm int},
\label{upper_bound_of_intractable_uncetainty}
\end{align}
for any number $n$, that is, the uncertainty in the average is also bounded by the parameter $\delta_{\rm int}$.

When the fluctuation of the output $n$ needs to be taken into account, we need to treat $\bar \epsilon_{\rm int}$ instead of $\epsilon_{\rm int}(n)$.
Fortunately, the averaged quantity continues to fulfill the same inequality as \eqref{upper_bound_of_intractable_uncetainty} as can be confirmed 
immediately,
\begin{align}
\label{intun2}
\bar \epsilon_{\rm int} 
&= |\tilde{X}_{N}-X_{N}| = \left\vert \sum_{n = 1}^N P(n)\, (\tilde{X}_{n}-X_{n}) \right\vert \nonumber \\
&\leq \sum_{n = 1}^N P(n)\, \left\vert \tilde{X}_{n}-X_{n}\right\vert  \leq \sum_{n = 0}^N P(n)\, \delta_{\rm int} = \delta_{\rm int}.
\end{align}

\subsubsection{Statistical Uncertainty}

The statistical error derives from the finiteness of the outputs obtained in the experiment, which may be specified by the assurance level we wish to grant for the result of the measurement.  

To this end, we invoke \emph{Chebychev's inequality} which provides a convenient evaluation by means of the variance of the random variable under consideration.  In the present case, it asserts that the cumulative distribution function
\begin{equation}\label{def:Tchebychev_inequality}
\mathrm{Pr}\left[ \epsilon_{\rm st}(n) \leq \delta_{\rm st}\right] \geq T(\delta_{\rm st};n)
\end{equation}
of the deviation to be less than a bound $\delta_{\rm st} > 0$ can be evaluated from below by the lower-bound function
\begin{equation}\label{def:Tchebychev_lower-bound}
T(\delta_{\rm st};n):= \max \left[ 1 - \frac{(\triangle x)^2} {n\delta_{\rm st}^{2}},\, 0 \right],
\end{equation}
with 
\begin{equation}\label{mesvar}
(\triangle x)^2 := \lim_{n \to \infty} \sum_i^n \frac{1}{n}(x_{i} - X_{\infty})^2
\end{equation}
being the variance of the distribution of the true values.

Now, let $\eta \in [0,1]$ be our assurance level of measurement, namely, the statistical uncertainty is assured to be smaller than a certain bound $\delta_{\rm st}$ 
with probability $\eta$.  From Chebychev's inequality \eqref{def:Tchebychev_inequality}, we can put $\eta = T(\delta_{\rm st};n)$, which may be used to determine
the bound $\delta_{\rm st}$ by solving the condition
\begin{equation}
\label{etakappa}
\eta = 1 - \frac{(\triangle x)^2}{n\delta_{\rm st}^{2}},
\end{equation}
unless $n$ is too small or 
$(\triangle x)^2$ is too large for which $T(\delta_{\rm st};n)$ vanishes.  We are thus left with the bound for the statistical uncertainty,
\begin{equation}
\epsilon_{\rm st} = \left|{X}_{n}-X_{\infty}\right| \leq \sqrt{\frac{(\triangle x)^2}{n(1-\eta)}}.
\end{equation}
Although we cannot know the variance $(\triangle x)^2$ given in \eqref{mesvar} without performing the observations infinite times, we may instead use the theoretical variance 
evaluated for the meter state $\ket{\zeta}$, namely,
\begin{equation}\label{varieq}
(\triangle x)^2 = \braket{\zeta|X^2|\zeta} - \braket{\zeta|X|\zeta}^2,
\end{equation}
which holds true when our theoretical model is exact (see the discussion of the model uncertainty below together with \eqref{modun}).

Now, to accommodate the case of fluctuation of output number $n$, we first introduce the function,
\begin{equation}\label{def:lower-bound_probability}
r(\delta_{\rm st};N) := \sum_{n=1}^{N} T(\delta_{\rm st};n) \, P(n),
\end{equation}
after multiplying the both sides of Chebychev inequality \eqref{def:Tchebychev_inequality}
by the probability $ P(n)$, and summing over $n$ from $0$ to $N$, 
we obtain
\begin{equation}\label{ineq:eval_of_CDF}
\mathrm{Pr}\left[ \bar \epsilon_{\rm st} \leq \delta_{\rm st}\right] \geq r(\delta_{\rm st};N).
\end{equation}

At this point, we observe that, for our case of the probability distribution being given by the binomial distribution \eqref{def:binom_distr}, 
the previous evaluation \eqref{def:Tchebychev_inequality} is a special case of \eqref{ineq:eval_of_CDF} for $q=1$.  It is intuitively clear that the lower-bound function \eqref{def:lower-bound_probability} is a monotonically increasing 
function with respect to all of its parameters $\delta_{\rm st}$, $N$ and $q$.  For a rigorous proof of the monotonicity, see Appendix~\ref{app:monotonicity}.

Employing a similar argument used for solving $\eta$ in favor of $\delta_{\rm st}$ in \eqref{etakappa}, thanks to the monotonicity of the function $r(\delta_{\rm st};N)$ we can solve $\eta = r(\delta_{\rm st};N)$ to obtain its solution $\delta_{\rm st}(\eta;N,q)$ which is then used to transform \eqref{ineq:eval_of_CDF} into 
\begin{equation}
\label{kappakun}
\bar \epsilon_{\rm st} \leq \delta_{\rm st}(\eta;N,q).
\end{equation}
This shows that on average the statistical error 
is guaranteed to be bounded from above with probability not less than $\eta$.
Due to the monotonicity of the lower-bound function \eqref{def:lower-bound_probability}, it is straightforward to see by definition that $\delta_{\rm st}(\eta;N,q)$ increases monotonically with respect to $\eta$, while it decreases monotonically with respect to $N$ and $q$.

\subsubsection{Model Uncertainty}

We now turn to the theoretical error given by \eqref{theoun}.
First, for the component of the model error $\epsilon_{\rm mod}$ in \eqref{modun}, as we mentioned earlier we regard our model to be exact enough so that the model error can be ignored safely,
$\epsilon_{\rm mod} = 0$.
In fact, the two experiments which we analyze later require fairly simple settings for which the theoretical modeling is reasonably easy, and
this allows us to assume that the upper bound of the model error $\epsilon_{\rm mod}$ can be neglected, that is, the model uncertainty $\delta_{\rm mod}$ is virtually zero $\delta_{\rm mod} = 0$ compared to the other uncertainties in our analysis.

\subsubsection{Approximation Uncertainty}

As for the component of the approximation error $\epsilon_{\rm app}$ in \eqref{appun}, we just consider the error that arises from our approximation
of using only the first order of the coupling parameter $\theta$ for the evaluation of the meter shift.  This is expected to provide a dominant factor for the error, given that the amplification of the weak value is expected to ruin the linear approximation when it is taken to be too large.
On account of this, and also combined with the fact that the full result can be obtained under the use of the Gaussian mode for the initial meter state in the analysis, in the present paper we assume that our approximation error just consists of the difference between the linear approximation 
\eqref{def:shift2} and the full result.  In short, we regard the uncertainty associated with the approximation to be given by the error itself,
\begin{align}
\label{apunclinear}
\delta_{\mathrm{app}} = \epsilon_{\mathrm{app}} = \left| \Delta_{\rm w}^{(1)}(\theta)-\Delta_{\rm w}(\theta)\right|.
\end{align}

\subsection{Relative Uncertainty}

Collecting all the uncertainties mentioned above, we find that the total operational difference is bounded by
\begin{align}
\bar \epsilon_{\mathrm{op}} \leq \Gamma,
\end{align}
with the total uncertainty,
\begin{equation}
\label{totalub}
\Gamma := \delta_{\rm int}+\delta_{\rm st}(\eta;N,q)+\delta_{\mathrm{app}}.
\end{equation}
We then introduce the {\it relative uncertainty} by the ratio of the shift in the actual experiment to the bound of operational difference:
\begin{align}
\label{relun}
R:=\frac{\Gamma}{\Delta_{\rm w}(\theta)}.
\end{align}
With this $R$, one may conclude that {\it the measurement is valid if $R < 1$ and invalid otherwise}.  A point to be noted is that for arguing the validity of measurement the ratio $R$ is more appropriate than the direct output values which are dependent on the scale of the measurement device.  

\section{\label{sec:examination}Examination of two preceding experiments}

We now examine, based on our theoretical framework, the two preceding experiments mentioned in the Introduction, that is, 
the USBD experiment by Dixon {\it et al}.~\cite{Dixon_2009} and the SHEL experiment by Hosten and Kwiat \cite{Hosten_2008}.

\subsection{\label{sec:Dixon}The USBD Experiment}

Dixon and his collaborators have shown that tiny beam deflections and their corresponding angular deflections of a mirror can be detected using weak values \cite{Dixon_2009}.  We first analyze this experiment to examine its validity as precision measurement based on our theoretical framework.

\subsubsection{Setup and Goal of the Experiment}

\begin{figure}
\centering
\includegraphics[width=8.5cm]{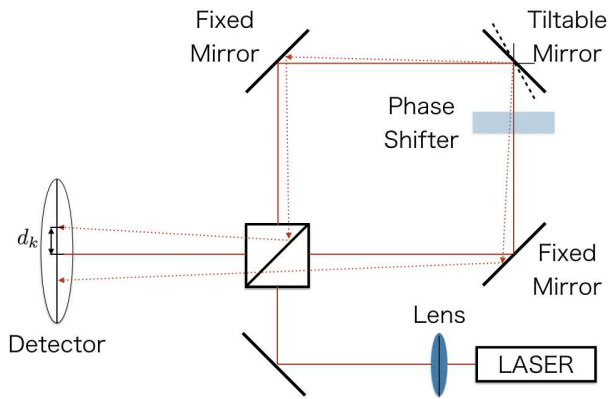}
\caption{The setup of the USBD experiment, with a slight modification for the convenience of our argument. 
After the photon passes the beam splitter, it goes either clockwise or counterclockwise along the path before entering the beam splitter again.
}\label{fig:sag}
\end{figure}

The setup of the experiment consists of a laser, a lens, a Sagnac interferometer and a position (quadrant) detector, where one of the mirrors in the interferometer is tiltable (Fig.\ref{fig:sag}).  The experimental group estimated the tiny angular deflection of the tiltable mirror from the beam deflection
 in the transverse direction obtained at the detector placed at the end of the path.
The benefit of the weak measurement is that by means of \lq amplification\rq\ the deflection of the transverse position on the detector can be much larger than the deflection $d_{k}$ expected from geometrical optics (which is shown by the incoming points of the dotted lines at the detector in Fig.\ref{fig:sag}).

The beam used in the experiment has the total wave number $k_0$, and 
when the tiltable mirror changes its angle, the path of the photon is slightly altered in the interferometer acquiring a small amount of transverse momentum $\hbar k$ in the clockwise case or $-\hbar k$ in the counterclockwise case.  
In addition, a 50/50 beam splitter is inserted together with a phase shifter, which consists of a half wave plate and a Soleil-Babinet compensator, in order to prepare both the preselected state and the postselected state in the experiment.  The amplification of the angle deflection of the mirror will then be examined by looking at the average position of photons arriving at the detector.   

To describe the system, we may use the single photon model as follows.  We separate the degrees of freedom of a photon into two parts, one associated with the freedom as to which path the photon takes after it exits the beam splitter, and another associated with the freedom in the transverse position of the photon along the path it takes.
The former degrees of freedom are considered to be the target system $\mathcal{H}$ whereas the latter corresponds to the auxiliary meter system $\mathcal{K} =L^{2}(\mathbb{R})$ where we choose $X = \hat x$ for the observable of the measurement.
Denoting the basis states in the former space $\mathcal{H}$ by $\ket{l}$ when it takes the counterclockwise path and similarly by $\ket{r}$ when it takes the clockwise path, we have
\begin{align}
\mathcal{H}&=\mathrm{span}\{\ket{l},\ket{r}\}.
\end{align}
In this experiment, the meter state $\ket{\chi_{0}}\in \mathcal{K}$ is prepared by the Gaussian state $\braket{x|\chi_{0}} = G(x)$ given by \eqref{def:Gaussian_function}.
To find the time development of the meter state, we employ the {\it paraxial approximation} with the effect of the lens being taken into account, 
in which the unitary operator $V_0$ needed to obtain the meter state $\ket{\chi}$ reads (see Appendix \ref{paraxial})
\begin{align}
\label{dixon_preop}
V_0 = e^{i\frac{l_{\mathrm{lm}}}{2k_0}\hat{k}^{2}_{x}}e^{i\frac{k_0}{2s_{\mathrm{i}}}\hat{x}^{2}},
\end{align}
where $l_{\mathrm{lm}}$ is the distance between the lens and the tiltable mirror along the beam path, $s_{\mathrm{i}}$ is the image distance of the lens \cite{koiketanaka_11} (see Fig.\ref{fig:propagation}), and $\hat k_x := \hat p_x /\hbar$ is the wave number operator along the $x$ direction (here we have used $\hat p_x$ for the momentum operator conjugate to $\hat x$ which we denoted by $\hat p$ earlier).
The initial meter state, given by \eqref{prechi}, now reads
\begin{align}
\label{prechi2}
\braket{x|\chi} = \left(\frac{\alpha}{2\pi(\alpha^2 + \beta^2)}\right)^{\frac{1}{4}}
\exp\left(-\frac{\alpha-i\beta}{4(\alpha^2 + \beta^2)}x^2\right)
\end{align}
where
\begin{align}
\alpha = \frac{s_{\mathrm{i}}^2 a^2}{s_{\mathrm{i}}^2 + 4k_0^2 a^4}, 
\qquad \beta = \frac{l_{\mathrm{lm}}}{2k_0} + \frac{2s_{\mathrm{i}}k_0 a^4}{s_{\mathrm{i}}^2 + 4k_0^2 a^4},
\end{align}
for which the average of the meter position vanishes $\braket{\chi | X |\chi} = 0$.

\begin{figure}
\centering
\includegraphics[width=6cm]{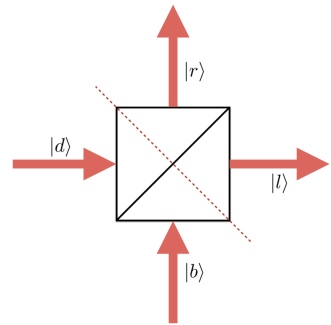}
\caption{Input/output relations of the beam splitter.}\label{fig:bs}
\end{figure}

To describe the process of the beam splitter, we consider a matrix $M_{\mathrm{BS}}$ corresponding to the process
\begin{align}
M_{\mathrm{BS}}=\frac{1}{\sqrt{2}}I +\frac{i}{\sqrt{2}}\left(\ket{l}\bra{r}+\ket{r}\bra{l}\right)
\end{align}
In the experiment, incoming photons come from below in Fig.\ref{fig:bs}, namely as the state $\ket{b}(=\ket{r})$ and outgoing photons are written as
\begin{align}
M_{\mathrm{BS}}\ket{b}&=M_{\mathrm{BS}}\ket{r}=\frac{1}{\sqrt{2}}(\ket{r}+i\ket{l}).
\end{align} 

\subsubsection{Measurement Operator and State Preparation}

In order to characterize the choice of the path of the photons, we may introduce  the observable operator
\begin{equation}
A=\ket{r}\bra{r}-\ket{l}\bra{l},
\label{eq: pathoperator}
\end{equation}
which has the eigenvalue $+1$ for the counterclockwise path and $-1$ for the clockwise path.
The unitary evolution operator describing the reflection of the beam at the tiltable mirror is
\begin{equation}\label{dixon_meas_op}
\begin{split}
U(k)&=e^{-ik A\otimes\hat{x}}.
\end{split}
\end{equation}
This unitary operator is the one we encountered as \eqref{def:unitary_interaction} in section \ref{sec:theoretical_framework}, where the parameter $\theta$ now corresponds to $\hbar k$, the target system operator $A$ is given by \eqref{eq: pathoperator}, and $Y$ is given by the position operator $\hat{x}$. 
Clearly, our theoretical model fits into the framework presented in the previous sections.

The preselected state in the present experiment is the state just before the reflection at the tiltable mirror. 
Note that,  the photons which take the counterclockwise path pass through the phase shifter before the reflection, while those which take the clockwise path do not (see Fig.\ref{fig:sag}).  Consequently, the phase shift operation on the photon which take the counterclockwise path reads 
\begin{align}
 O_r&=e^{i\frac{\varphi}{2}}\ket{r}\bra{r}+\ket{l}\bra{l}\label{phase_shifter_l}.
\end{align}
where the phase $\varphi /2$ is provided by the phase shifter. The preselected state is given by
\begin{align}
\ket{\psi}&=O_rM_{\mathrm{BS}}\ket{b}\label{init_Dix} \\
&=\frac{1}{\sqrt{2}}\left(ie^{i\frac{\varphi}{2}}\ket{r}+\ket{l}\right).
\end{align}
On the other hand, the postselected state $\ket{\phi}$ corresponds to the state just after the reflection on the tiltable mirror is given by
\begin{align}
\ket{\phi}&=O_rM_{\mathrm{BS}}\ket{d}\label{eq:finst}\\
&=\frac{1}{\sqrt{2}}\left(\ket{r}+ie^{-i\frac{\varphi}{2}}\ket{l}\right),
\end{align}
where the phase shift operation to the photon which goes clockwise reads
\begin{align}
O_l=\ket{r}\bra{r}+e^{i\frac{\varphi}{2}}\ket{l}\bra{l}.
\end{align}
We mention that, although our expressions of the preselected state and the postselected state are slightly different from the original ones given in \cite{Dixon_2009}, the outcome of the measurement remains the same as shown in the Appendix.

\begin{figure}
\centering
\includegraphics[width=7cm]{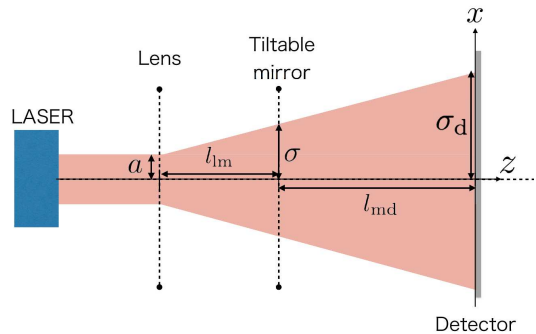}
\caption{Extension of the width of the Gaussian state of the meter by the lens and propagation.  The original width $a$ at the laser is extended to $\sigma$ at the tiltable mirror and further extended to $\sigma_{\mathrm{d}}$ at the detector.  Here we have made the entire path of the photons into a straight line so that the extension of the width along the passage from the laser to the detector is readily seen.}
\label{fig:propagation}
\end{figure}

The weak value, which is defined in \eqref{def:weakvalue}, then reads
\begin{equation}
\label{eq:wv}
\begin{split}
A_{\mathrm{w}} = i\cot{\frac{\varphi}{2}},
\end{split}
\end{equation}
which is purely imaginary.
Note that, for small $\varphi$, the transition amplitude $\vert {\braket{\phi|\psi}\vert^2}$ becomes small as well, implying that the beam coming out of the Sagnac interferometer toward the detector will be dark.
In other words, we employ this \lq dark port\rq\ in our weak measurement by choosing the dark beam for the postselected state.

\begin{figure*}[t]
\centering
\includegraphics[width=16cm]{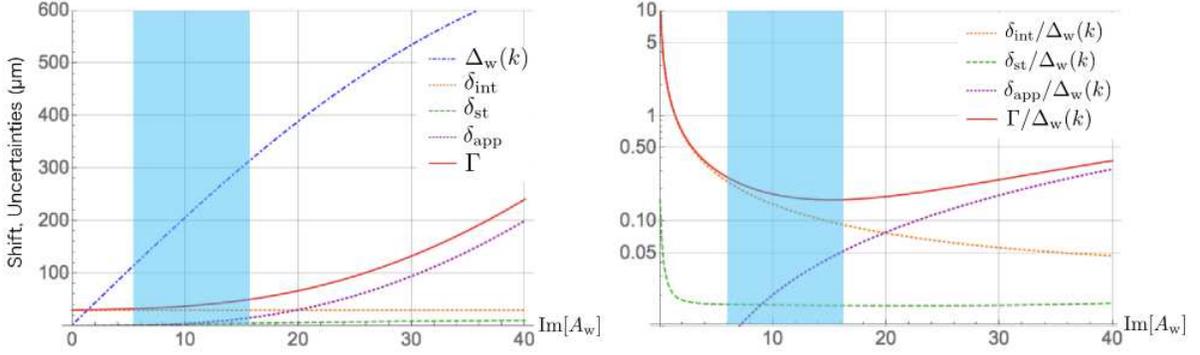}
\caption{Shift/uncertainties (left) and relative uncertainties (right) as functions of ${\rm Im}[A_{\rm w}]$ obtained at $\sigma_{\mathrm{d}} = 750\mu\mathrm{m}$.  The zones colored in cyan cover the actual values of ${\rm Im}[A_{\rm w}]$ used in the experiment.}
\label{UVR750}
\end{figure*}

After the postselection, we obtain the meter state $\ket{\xi}$ in 
\eqref{postselst}, {\it i.e.},
\begin{align}
\ket{\xi} = \frac{\bra{\phi}e^{-ik A\otimes\hat{x}}\ket{\psi} |\chi\rangle}{||\bra{\phi}e^{-ik A\otimes\hat{x}}\ket{\psi}\ket{\chi}||}.
\end{align}
Here we need the unitary operator $V_1$ that represents the time development  of the meter state $\ket{\xi}$ under the beam propagation from $z=0$ to $z=l_{\mathrm{md}}$ before the photon reaches the detector.
In our paraxial approximation (see Appendix \ref{paraxial}), we have
\begin{equation}
\label{vone}
V_1=e^{-i\frac{\hat{k}^{2}_{x}}{2k_0}l_{\mathrm{md}}},
\end{equation}
which yields the meter state on the detector $\ket{\zeta} = V_1 \ket{\xi}$ in \eqref{zeta}.
Adopting the procedure given in \cite{koiketanaka_11}, one can obtain
the expectation value of the meter observable $\ket{x}$ as
\begin{align}
\braket{\zeta| \hat x |\zeta} =\frac{2k\,{\rm Im}[A_{\rm w}]\sigma_{\mathrm{d}} \sigma\exp[-2k^{2}\sigma^{2}]}{1+\frac{1}{2}({\rm Im}[A_{\rm w}]^{2}-1)(1-\exp[-2k^{2}\sigma^{2}])},
\label{eq:Dixon_fullorder_displ}
\end{align}
where 
\begin{equation}
\sigma=\frac{l_{\mathrm{md}}a +l_{\mathrm{lm}}\sigma_{\mathrm{d}}}{l_{\mathrm{lm}}+l_{\mathrm{md}}},
\end{equation}
which is defined from the path length $l_{\mathrm{lm}}$ between the lens and the mirror and also from the path length $l_{\mathrm{md}}$ between the mirror and the detector.  In the above, $a=640\mathrm{\mu m}$ is the beam width at the lens provided by the laser beam source (see Fig.\ref{fig:propagation}) and $\sigma_\mathrm{d}$ is the beam width at the detector tuned by the lens.
In the linear approximation in the coupling $k$, the result \eqref{eq:Dixon_fullorder_displ} reduces to
\begin{align}
\braket{\zeta| \hat x |\zeta} = 2k\,{\rm Im}[A_{\rm w}]\sigma_{\mathrm{d}} \sigma+O(k^{2}).
\label{eq:Dixon_1storder-displ}
\end{align}

We note that with the unitary operator $V_1$ in \eqref{vone} we have 
\begin{equation}
\label{eq:developed_x}
\braket{\chi |V^{\dag}_1 X V_1|\chi} = 0
\end{equation}
so that the condition \eqref{chicond} is fulfilled.
From this, we learn that the full shift $\Delta_{\rm w}(k)$ corresponding to \eqref{def:shift} (where we have set $\theta = \hbar k$ as mentioned before) is given by $\Delta_{\rm w}(k) = \braket{\zeta| \hat x |\zeta}$ in \eqref{eq:Dixon_fullorder_displ}, 
while that of the linear approximation \eqref{def:shift2} is given by
\begin{align}
\label{deltalin}
\Delta_{\rm w}^{(1)}(k) = 2k\,{\rm Im}[A_{\rm w}]\sigma_{\mathrm{d}} \sigma.
\end{align}
The approximation error $\epsilon_{\mathrm{app}}$, which is equal to the approximation uncertainty $\delta_{\mathrm{app}}$ in our approach, can be obtained from the difference between these shifts.

\subsubsection{Parameter Setting for Simulation and Results}

To estimate the uncertainties $\delta_{\rm int}, \delta_{\rm st}(\eta;N,q), \delta_{\mathrm{app}}$ explicitly, we need to know the parameters used in the actual experiment.
First, to determine the value of $\delta_{\rm int}$,  we recall that in the experiment the measurement outcome is shown with \lq random error\rq, which presumably refers to the statistical deviation with respect to different bunches of photons injected in every 1/200 second.  This allows us to choose the value 
\begin{align}
\delta_{\rm int} = 30\ \mathrm{\mu m}, 
\end{align}
which is the maximal size of error bar in the experiment data.

Next, the beam used has the total 
wave number  $k_0 = {2\pi}/{\lambda}=8.06\, \mu \mbox{m}^{-1}$, which corresponds to the wave length $\lambda = 780\, {\rm nm}$.
The angle of the tilted mirror is fixed in such a way that the unamplified deflection at the detector is $d_{k} = 2.95\, \mu\mbox{m}$ (see Fig.\ref{fig:sag}), which translates into the value $k=2.08\times 10^{-5} \mu\mbox{m}^{-1}$.
Then, the trial number $N$, which is the number of photons injected in the interferometer, may be estimated from the laser power
$P = 3.2\ \mathrm{mW}$ and the wave length $\lambda$.
From these values, the 
photon number per second is estimated as
\begin{align}
\frac{P\lambda}{hc}=1.3\times 10^{16}\ \mathrm{sec}^{-1},
\end{align}
where $h$ is the Planck constant and $c$ is the speed of light.  Since the 
sampling rate of the detector is $200\ \mathrm{Hz}$,  the photon number $N$ is found to be
\begin{align}
\label{numbern}
N=\frac{P\lambda}{hc}\cdot\frac{1}{200}=6.5\times 10^{13}.
\end{align}
In our numerical analysis, however, we adopt the number $10^{8}$ for $N$ for computational convenience. 
This is allowed because one can show that a smaller $N$ gives a larger statistical uncertainty (see Appendix \ref{app:monotonicity}) and hence
our examination provides a stricter condition than the actual one. 
Another point that should be noted in this regard is that, 
in the actual experiment, an extra 50/50 beam splitter is inserted to observe the beam structure with a CCD camera (which are not shown in Fig.\ref{fig:sag}).  Although this beam splitter will halve the number of photons that reach the detecter, this change is negligible since it is much smaller than the ratio of reduction from $10^{13}$ to $10^{8}$ mentioned above.

\begin{figure*}[t]
\centering
\includegraphics[width=16cm]{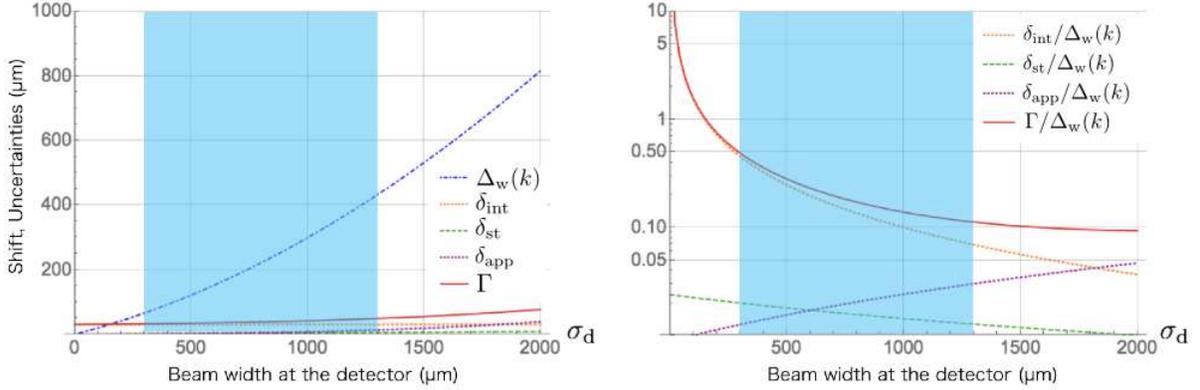}
\caption{Shift and various uncertainties (left), the relative uncertainty $R = \Gamma/\Delta_{\rm w}$ and its components (right) as functions of $\sigma_{\mathrm{d}}$ obtained at ${\rm Im}[A_{\rm w}]=10$.
The zones colored in cyan cover the actual values of $\sigma_{\mathrm{d}}$ used in the experiment.}
\label{width-graph}
\end{figure*}

In order to find out the value of $\delta_{\rm st}(\eta;N,q)$, we need to choose our assurance level $\eta$ of measurement.  We do this by setting $\eta = 0.95$ as a reasonable level 
of statistical confidence.  Another ingredient needed is the quadratic moment of the meter state $\ket{\zeta}$,
\begin{align}
&\braket{\zeta| \hat x^2 |\zeta} \nonumber \\
&= \frac{\sigma_{d}^{2}\left(1+{\rm Im}[A_{\rm w}]^2+(1-{\rm Im}[A_{\rm w}]^2)(1-4k^2\sigma^2)e^{-2k^2\sigma^2}\right)}{1+{\rm Im}[A_{\rm w}]^2+(1-{\rm Im}[A_{\rm w}]^2)e^{-2k^2\sigma^2}}
\end{align}
which is used to evaluate the lower-bound function $T(\delta_{\rm st};n)$ in \eqref{def:Tchebychev_lower-bound} by estimating the variance $(\triangle x)^2$ in \eqref{mesvar}
based on the relation \eqref{varieq} for $X = \hat x$.
With the value of $N$ in \eqref{numbern} and the postselection rate $q$ in \eqref{postselcrate},  the value of $\delta_{\rm st}(\eta;N,q)$ can be determined numerically.

These two upper bounds obtained above, $\delta_{\rm int}$, $\delta_{\rm st}(\eta;N,q)$, are shown in Fig.\ref{UVR750} (left) as functions of ${\rm Im}[A_{\rm w}]$, together with 
$\Delta_{\rm w}(k)$, $\epsilon_{\mathrm{app}}$, for the width $\sigma_{\mathrm{d}} = 750\ \mathrm{\mu m}$ which we choose as a representative point.   
The the total upper bound $\Gamma\ = \delta_{\rm int}+\delta_{\rm st}(\eta;N,q)+\delta_{\mathrm{app}}$ in \eqref{totalub} is also indicated.
The curves shown in Fig.\ref{UVR750} (right) are relative ratios for each of the uncertainties with respect to the shift and their total ratio which is the relative uncertainty $R$ in \eqref{relun}.    

By looking at the shift formula \eqref{eq:Dixon_fullorder_displ} and \eqref{eq:developed_x}, we observe that 
the amplification is implemented in the experiment by two means: one through the choice of the relative phase $\varphi$ which leads to the amplification of the weak value $A_{\mathrm{w}}$ (or more precisely its imaginary part ${\rm Im}[A_{\rm w}]$) itself, and the other through the choice of the width $\sigma_{\mathrm{d}}$ of the Gaussian mode of the meter state.  In what follows, the validity of the measurement under each of these two means is examined separately.

\paragraph{Weak Value Dependence.}
In the experiment, the phase $\varphi$ was chosen so that the imaginary part of the weak value ${\rm Im}[A_{\rm w}]$ in \eqref{eq:wv} becomes $6.57,\ 9.93,\ 15.9$.  
In Fig.\ref{UVR750},
the zone colored in cyan covers the three values of ${\rm Im}[A_{\rm w}]$ used in the experiment. 
Viewed from the shift $\Delta_{\rm w}(k)$ and the relative uncertanty $R$, we immediately observe that, according to our criterion mentioned when we defined $R$ before, the three values ${\rm Im}[A_{\rm w}]$ used in the experiment are in the safe region where the measurement is valid.  
In fact, it is amusing to observe that the ratio $R$ has its minima at around ${\rm Im}[A_{\rm w}] \approx 15$ and that the largest value ${\rm Im}[A_{\rm w}] = 15.9$ used in the experiment is
found to be close to this optimal point.

\paragraph{Beam Width Dependence.}
In the experiment, in addition to the value $\sigma_{\mathrm{d}} = 750\ \mathrm{\mu m}$, various other 
values in the region $300 < \sigma_{\mathrm{d}} < 1300$ are also adopted for the beam width.  
We analyze the variation of the shift $\Delta_{\rm w}(k)$ and the relative uncertainty $R$ when we vary the width $\sigma_{\mathrm{d}}$ 
at the fixed value of the weak value ${\rm Im}[A_{\rm w}]=10$ (see Fig.\ref{width-graph}).  As before, for both the shift $\Delta_{\rm w}(k)$ and the relative uncertainty $R$, the operational difference is determined primarily by the intractable uncertainty.   Our analysis shows that using larger beam widths can improve the ratio $R$, if the technical problems which have prevented us from going there in the actual experiment can be removed.

\subsection{\label{sec:Hosten}The SHEL Experiment}

Hosten and Kwiat observed the spin Hall effect using weak measurement \cite{Hosten_2008} by detecting an extremely tiny spin-dependent shift perpendicular to the refractive index gradient for photons.  We next analyze this measurement to examine if it is valid as precision measurement according to our criterion.

\subsubsection{Setup and Goal of the Experiment}
\begin{figure}[t]
\centering
\includegraphics[width=8cm]{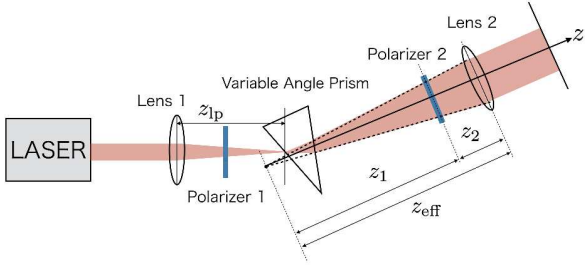}
\caption{Schematic diagram of the SHEL experiment.  Here, $z_{\rm eff} = z_1 + z_2$ is the effective focal length of Lens 2.}
\label{hosten-experiment}
\end{figure}

The setup consists of a laser, two polarizers, two lenses, a variable angle prism, a half wave plate, and a position detector (Fig.\ref{hosten-experiment}).
When the laser beam passes through the prism, spin-dependent shift perpendicular to the incident direction occurs on the variable angle prism.
This spin Hall effect predicts position shifts of photons, which are too tiny to detect with a simple use of the photon detector.   With weak
measurement, Hosten and Kwiat has succeeded to detect it by amplifying the shift considerably.  

Analogously to the previous case, the experiment is modeled by separating the degrees of freedom of a photon into two parts, one associated with the spin freedom, and another associated with the freedom in the transverse position of the photon along the path it takes. 
The former degrees of freedom are considered to be the target system $\mathcal{H}$ whereas the latter corresponds to the auxiliary meter system $\mathcal{K}$. 
Denoting the basis states in $\mathcal{H}$ by $\ket{+}$ when the spin is along the direction of the propagation while by $\ket{-}$ when it is along the opposite direction, 
we have
\begin{align}
\mathcal{H}=\mathrm{span}\{\ket{+},\ket{-}\},
\end{align}
again with $\mathcal{K}=L^{2}(\mathbb{R})$.

The initial state $\chi(x)$ of the meter system is obtained by 
taking account of the propagation effect and the lens effect during the passage from Lens 1,  where
we prepare the Gaussian state $\braket{x|\chi_{0}} = G(x)$ in \eqref{def:Gaussian_function}, 
to the prism.  This is done, as in the previous case \eqref{dixon_preop},
by applying the unitary operator,
\begin{align}
V_0 = e^{i\frac{z_{\mathrm{lp}}}{2k_{0}}\hat{k}^{2}_{x}}e^{-i\frac{k_{0}}{2z_{\mathrm{lp}}}\hat{x}^{2}},
\end{align}
on the state $\ket{\chi_{0}}$, where $z_{\mathrm{lp}}$ is the distance between Lens 1 and the prism (see Fig.\ref{hosten-experiment}).

The initial meter state, given by \eqref{prechi}, takes again the form \eqref{prechi2} but with
\begin{align}
\alpha = \frac{z_{\mathrm{lp}}^{2} a^{2}}{z_{\mathrm{lp}}^2 + 4k_{0}^{2} a^{4}}, 
\qquad \beta = \frac{-z_{\mathrm{lp}}}{2k_0} + \frac{2z_{\mathrm{lp}}k_0 a^4}{z_{\mathrm{lp}}^2 + 4k_0^2 a^4}.
\end{align}
In this case, we can easily check $z_{\rm lp}^{2}\ll k_{0}^{2}a^{4}$ with parameters given later (see the part of parameter setting below) so that 
$\beta$ can be ignored compared to $\alpha$.   This allows us to simplify \eqref{prechi2} into the Gaussian function $G(x)$ with $a$ replaced by
$\alpha$, that is, 
\begin{align}
\braket{x|\chi} = \left(\frac{1}{2\pi\alpha}\right)^{\frac{1}{4}}\exp\left(-\frac{x^{2}}{4\alpha}\right).
\end{align}

\subsubsection{Measurement Operator and State Preparation}
In order to characterize the spin, we may introduce the operator 
\begin{equation}
A=\ket{+}\bra{+}-\ket{-}\bra{-},
\label{A_hosten}
\end{equation}
which has the eigenvalue $+1$ for the $\ket{+}$ state and $-1$ for the $\ket{-}$ state. When the beam goes through the prism, the beam is slightly shifted in the direction perpendicular to the direction of beam propagation with the sign $\pm$ depending on the spin $\ket{\pm}$ of the photon, which is the spin Hall effect. 
Introducing the coordinate in which the beam propagates along the $z$-direction and the shift takes place in the $x$-direction, 
the unitary evolution operator describing this is found to be
\begin{align}
U_{g}=e^{-igA\otimes\hat{k}_{x}}.
\label{meas_op_hosten}
\end{align}
Here, the parameter $\theta$ in \eqref{def:unitary_interaction} in section \ref{sec:theoretical_framework} now corresponds to $g$ which is determined by the shift, and the operator $A$ is given by \eqref{A_hosten}.  The operator $Y$ corresponds to the momentum operator $\hat{k}_x$ perpendicular to  the direction of propagation.

\begin{figure}[t]
\centering
\includegraphics[height=7cm]{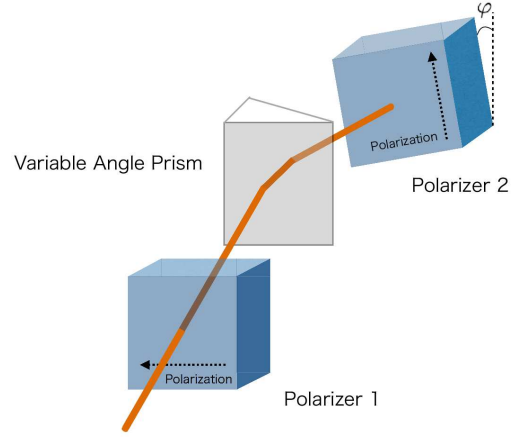}
\caption{Preselection by Polarizer 1 and postselection by Polarizer 2 tilted by angle $\varphi$.}
\label{selection_of_hosten-experiment}
\end{figure}

The preselected state $\ket{\psi}$ of the target system in the present experiment is 
\begin{equation}
\ket{\psi}=\frac{1}{\sqrt{2}}(\ket{+}+\ket{-}),
\end{equation}
which is prepared by adjusting the polarizer 1 properly.  We then choose the postselected state 
\begin{align}
\ket{\phi}=\frac{1}{\sqrt{2}i}\left(e^{i\varphi}\ket{+}-e^{-i\varphi}\ket{-}\right),
\end{align}
which can be realized by adjusting the angle parameter $\varphi$ of the polarizer 2 (see Fig.\ref{selection_of_hosten-experiment}).
The weak value, which is defined in \eqref{def:weakvalue}, then reads
\begin{align}
A_{\mathrm{w}}&=i\cot{\varphi}.
\end{align}
After the postselection, we obtain the meter state $\ket{\xi}$ in 
\eqref{postselst}, {\it i.e.},
\begin{align}
\ket{\xi} = \frac{\bra{\phi}e^{-igA\otimes\hat{k}_{x}}\ket{\psi} |\chi\rangle}{||\bra{\phi}e^{-igA\otimes\hat{k}_{x}}\ket{\psi}\ket{\chi}||}.
\end{align}

\begin{figure*}[t]
\centering
\includegraphics[width=16cm]{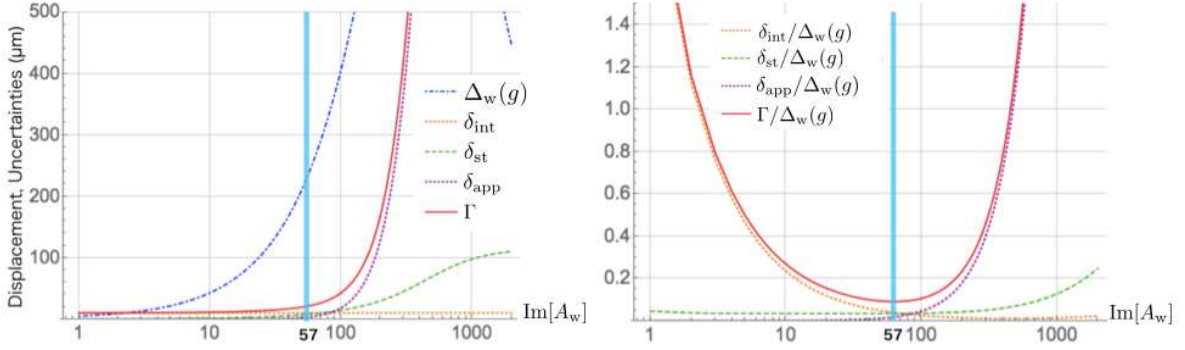}
\vspace{-.5cm}
\caption{Shift and various uncertainties (left), the relative uncertainty $R = \Gamma/\Delta_{\rm w}$ and its components (right) as functions of ${\rm Im}[A_{\rm w}]$, obtained at $g=27 \mathrm{nm}$ for which the angle of the prism smaller than $56^\circ$.   In this case, the actual value used in the experiment is ${\rm Im}[A_{\rm w}] = 57.3 \pm 0.7$ shown by the narrow strips colored in cyan.}
\label{wvdep}
\end{figure*}

To describe the meter state $\ket{\zeta}$, we adopt the paraxial approximation which we used in analyzing the USBD experiment. The change of the meter state after the beam enters the variable angle prism occurs in two steps.  The first is the change caused by the spin Hall effect implemented by the unitary operator \eqref{meas_op_hosten} we have already mentioned.  
Now we consider the meter state just after Lens 2 at $z_{\mathrm{eff}} = z_1 + z_2 \approx125 \mathrm{mm}$ (see Fig.\ref{hosten-experiment}).
As in the previous case of the experiment, we have
\begin{equation}
\label{vone2}
V_1=e^{-i\frac{\hat{k}^{2}_{x}}{2k_0}z_{\mathrm{eff}}},
\end{equation}
which yields the meter state on the detector $\ket{\zeta} = V_1 \ket{\xi}$ in \eqref{zeta}.

To calculate the position expectation value of this meter state, which corresponds to
the shift of the center of arriving photons yielding the signal, one can use \eqref{Heisenberg-position} to obtain
\begin{align}
\braket{\zeta|\hat{x}|\zeta}=gF \cot{\varphi}+O(g^{2}),
\label{expx}
\end{align}
where $F := z_{\mathrm{eff}}/(2k_{0}\alpha)$ is the geometrical amplification factor.
In fact, in the present case where the meter state $\ket{\chi}$ is given by the Gaussian state, 
the expectation value can be evaluated fully without approximation \cite{Cho_2010, Lee_2014}, which yields the shift of the meter variable as
\begin{equation}\label{eq:gtorealx}
\braket{\zeta|\hat{x}|\zeta}
=\frac{gF{\rm Im}[A_{\rm w}]\exp[-\frac{g^{2}}{2\alpha}]}{1+\frac{1}{2}\left(({\rm Im}[A_{\rm w}])^{2}-1\right)\left(1-\exp[-\frac{g^{2}}{2\alpha}]\right)},
\end{equation}
where now we have ${\rm Im}[A_{\rm w}] = \cot{\varphi}$.  Obviously, for small $g \ll \sqrt{\alpha}$, the expectation value reduces to \eqref{expx}.

\begin{figure*}[t]
\centering
\includegraphics[width=16cm]{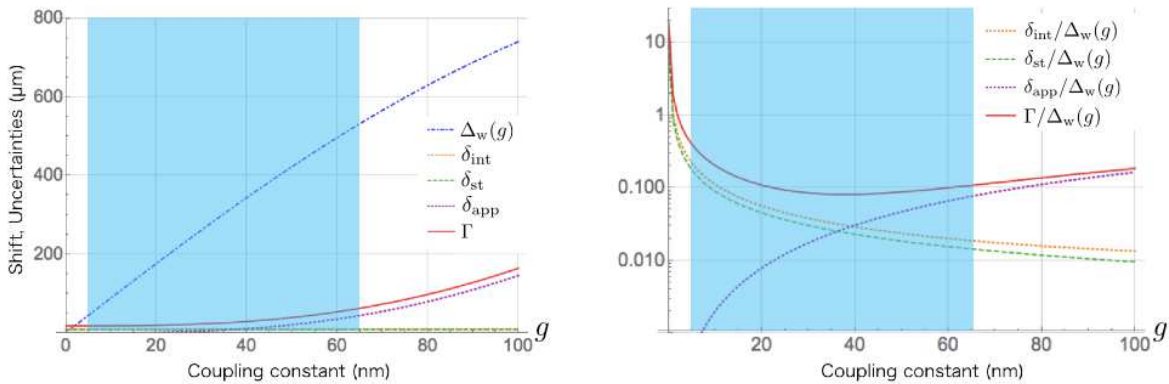}
\caption{Shift and various uncertainties (left), the relative uncertainty $R = \Gamma/\Delta_{\rm w}$ and its components (right) as functions of the coupling constant $g$, obtained at ${\rm Im}[A_{\rm w}]=57$.
The zones colored in cyan indicate the range of $g$ used in the experiment.  The relative uncertainty $R$ attains the minimum at around the value $g = 40$.}
\label{coupdep}
\end{figure*}

As in the previous case,  the shift of the meter variable is then given by the expectation value $\Delta_{\rm w}(g) = \braket{\zeta| \hat x |\zeta}$ in \eqref{eq:gtorealx}, while that of the linear approximation \eqref{def:shift2} is given by
\begin{align}
\label{deltalin2}
\Delta_{\rm w}^{(1)}(g) = gF \cot{\varphi}.
\end{align}
The approximation uncertainty $\delta_{\mathrm{app}} = \epsilon_{\mathrm{app}}$ can be obtained from the difference between these shifts.

\subsubsection{Parameter Setting for Simulation and Results}
As before, to estimate the uncertainties $\delta_{\rm int}, \delta_{\rm st}(\eta;N,q), \delta_{\mathrm{app}}$ explicitly, we need to figure out the parameters used in the actual experiment.

First, to determine the value of $\delta_{\rm int}$, we refer to Ref.\cite{Hosten_2008}. It is stated that the unwanted shifts caused by rotating the polarizer are about the order of 10$\mu {\rm m}$, which appears to provide a dominant factor to determine the intractable uncertainty in our discussion.  From this we choose our parameter, 
\begin{equation}
\delta_{\rm int} = 10\ \mu {\rm m}.
\end{equation}

As for the number of photons used in the experiment, we may estimate it as 
$3.2\times 10^{16}$ per second from the laser power $10\ \mathrm{mW}$ and the wave length $633\ \mathrm{nm}$ stated in \cite{Hosten_2008}.
However, in our numerical simulation, for the trial number $N$ we adopt the value $10^{9}$ for computational convenience, since the difference between them
is insignificant statisitically and does not affect the results of our simulation.

In order to find out the value of $\delta_{\rm st}(\eta;N,q)$, we again choose our assurance level $\eta$ by $\eta=0.95$. 
Another ingredient needed is the quadratic moment,
\begin{align}
&\braket{\zeta|\hat{x}^2|\zeta}\nonumber\\
&=\alpha+\frac{g^{2}\left(1+\frac{1}{2}({\rm Im}[A_{\rm w}]^{2}-1)\right)}{1+\frac{1}{2}({\rm Im}[A_{\rm w}]^{2}-1)(1-\exp[-\frac{g^2}{2\alpha}])}\nonumber \\
&\quad+\frac{z_{\rm eff}^{2}}{4\alpha k_{0}^{2}} \left(1+\frac{g^2}{\alpha}\frac{\frac{1}{2}({\rm Im}[A_{\rm w}]^{2}-1)\exp[-\frac{g^2}{2\alpha}]}{1+\frac{1}{2}({\rm Im}[A_{\rm w}]^{2}-1)(1-\exp[-\frac{g^2}{2\alpha}])}\right),
\end{align}
which is used to evaluate the lower-bound function $T(\delta_{\rm st};n)$ in \eqref{def:Tchebychev_lower-bound} by estimating the variance $(\triangle x)^{2}$ in \eqref{mesvar} based on the relation \eqref{varieq} for $X = \hat x$.  Again, 
with the value of $N$ and the postselection rate $q$ in \eqref{postselcrate}, the value of $\delta_{\rm st}(\eta;N,q)$ can be determined numerically.

In this experiment, the coupling $g$ depends on the setting of the variable angle prism, namely, the incident angle of the beam.  The main result
of the paper \cite{Hosten_2008} is the confirmation of the dependence predicted theoretically when the spin Hall effect exists.  
We check the validity of this experiment in view of two points, the choice of the weak value (or postselection) and the range of the coupling employed.

\paragraph{The Choice of the Weak Value.}
In the experiment, the postselection is made so that ${\rm Im}[A_{\rm w}]$ is fixed to the value $57.3 \pm 0.7$ for the angle of the prism smaller than $56^\circ$, or to the value $31.8 \pm 0.2$ for the angle of the prism larger than $56^\circ$.   
In both cases, the ratio $R$ is found well below 1, indicating that the measurement is valid (see Fig.\ref{wvdep}).  
It is noteworthy that the value ${\rm Im}[A_{\rm w}] = 57.3 \pm 0.7$ is close to the optimal point 64.2 of the ratio $R$.  

\paragraph{Coupling Dependence.}
In the experiment, the coupling constant is varied in the region $2 < g < 65$.  
We analyze the variation of the shift and uncertainties (Fig.\ref{coupdep}; left) and the relative uncertainties (Fig.\ref{coupdep}; right) in this region for $g$ at the fixed value of the weak value ${\rm Im}[A_{\rm w}]=57$.  From this, it is observed that the entire region of $g$ is in the safe zone of $R$ of less than 1, and that it covers the optimal value of $g$ around 40.   We thus find that the measurement as a whole is not just valid as prevision measurement but is almost an optimal one, even though the measurements with values near $g = 2$ are almost at the lower limit according to our criterion.  

\section{\label{sec:conclusion}Conclusion and Discussions}

In this paper, we have presented a general framework of evaluating the validity of
a weak measurement based on the analysis of uncertainty estimation associated with the measurement as well as the theoretical evaluation we use.  
The difference between the experimentally obtained value (the average shift of the meter) and the theoretically evaluated value is caused from the error in experiment and also from the error inherent to theory.  The experimental error arises from the statistical and the other intractable errors including the systematic ones, whereas the theoretical error originates from the modeling inaccuracy of the system and the error in approximation.   

These four types of errors are then bounded from above by their respective uncertainties determined from the experimental settings and the theoretical procedures.  
All these uncertainties are combined to furnish the total uncertainty associated with the weak measurement, and the ratio of the total uncertainty and the shift of the meter forms the relative uncertainty $R$ in \eqref{relun}.  Finally, the value of $R$ is used to determine the validity of the weak measurement:  if $R <1$, then the measurement is valid, while if $R \geq 1$, the measurement is invalid.

In order to evaluate the four uncertainties explicitly, we need to look into the detail of the actual procedures of the measurement and the theoretical calculation.  We may also be required to supplement this with a set of reasonable assumptions when no definite procedure to determine the uncertainties is available.  In the present paper, with the purpose of applying our framework to examine the two preceding experiments of weak measurement used for precision measurement, the SHEL experiment \cite{Hosten_2008} and the USBD experiment \cite{Dixon_2009}, we have adopted the following assumptions: 

\begin{enumerate}
\item The intractable uncertainty has a definite upper bound.
\item The statistical uncertainty stems solely from the probability distribution of the success rate of the postselection in the weak measurement.
\item The model we are using is accurate enough so that the model uncertainty can be ignored.  
\item The Gaussian state \eqref{def:Gaussian_function} prepared for the meter is exact.  This implies that our approximation uncertainty is just the difference between the results obtained by the linear approximation in the coupling constant and that of the full computation available for the Gaussian state.  
\end{enumerate}

\begin{figure*}[t]
\setlength\textfloatsep{300pt}
\centering
\includegraphics[width=16cm]{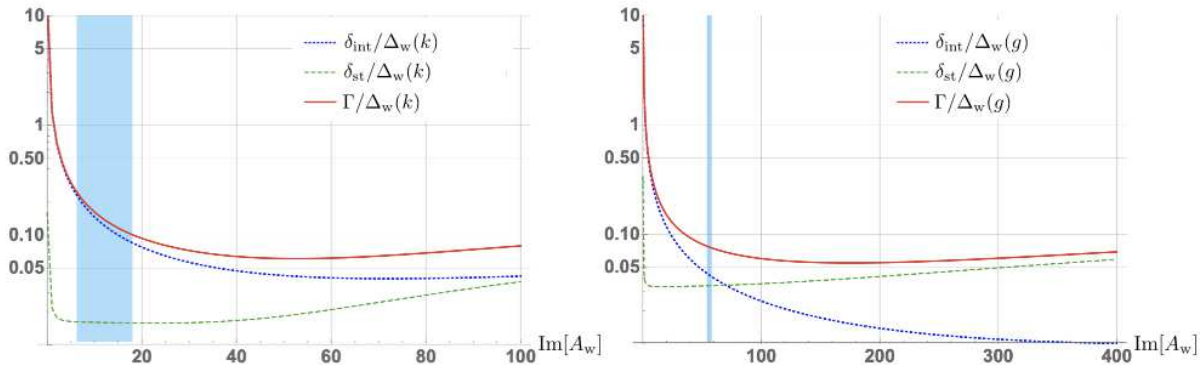}
\caption{Relative uncertainty $R = \Gamma/\Delta_{\rm w}$ and its components based on the full result of the meter shift for which $\epsilon_{\rm app} = 0$: 
the USBD experimente (left) and the SHEL experiment (right).}
\label{Fig:noapp}
\end{figure*}

Based on our scheme and the above supplemental assumptions, we have analyzed the aforementioned two experiments.  Our results show that both of the experiments are indeed valid as weak measurement, on the ground that all the scales of amplification of the weak value utilized for the precision measurement are well within the valid range of $R <1$.  
Moreover, it is observed that the scales of amplification used in the two experiments lay in the vicinity of the optimal point where $R$ takes its minimum.  

In our analysis, we have assumed that our approximation uncertainty is given by the linear approximation.  The reason for this is that, in general, the amplified meter shift is expected to be proportional to the weak value in the linear approximation, which has been the prime motivation for the application of weak measurement for precision measurement.  However, if the full result of the meter shift can be obtained without such approximation, we no longer need to
fall back on the evaluation of the approximation uncertainty with the linear approximation.  In fact, the preparation of the Gaussian state for the meter system considered in our argument offers such a case where the full result is available.

If we take this particularity into account,  we are allowed to set simply $\epsilon_{\rm app} = 0$.  Then, 
the outcome of the analysis, performed by eliminating the approximation uncertainty, is shown in Fig.\ref{Fig:noapp}.
This shows that, in principle, much stronger amplification of the weak value can be utilized than those used in the actual experiments attaining higher precision.  It should be noted, however, that in these experiments there might be causes which hampered the amplification to the extent of the desired level, due possibly to the ambiguity of the prepared meter state which could have different profiles from the the presumed Gaussian state, not to mention that there could be 
some technical noise that inevitably arises in the measurement to stifle the amplification.  For instance, the stray light incident mentioned in \cite{Dixon_2009} may be one of such technical sources of noise. 

We should also note that the result of our analysis may turn out to be quite different if a distinct assumption for the nature of the intractable uncertainty is adopted.  The ambiguity in the choice of probability distribution for the meter position provides another factor for obtaining different outcomes.
Besides, although we have chosen our intractable uncertainty to be independent of the trial number $N$, when the measurement outcome suffers from an external noise dictated by some probability distribution like those analyzed in \cite{Jordan_2014}, the intractable uncertainty becomes dependent on the trial number $N$, which could alter our result of analysis as well.  In any case, the determination of the intractable uncertainty seems to be an important element for improving our framework to obtain a more solid criteria for the validity of measurement in general.  

Despite all these technical details and ambiguities associated with the actual procedures of experiment and theoretical analysis, we hope that the scheme we presented in this paper serves as a basis to examine the validity of weak measurement in general, as such a method is very much in need in view of the future extension of the application of weak measurement, irrespective of whether it is for precision measurement or not.  
Our analysis on the two previous experiments given here may be regarded as a first test for the soundness of our scheme, and although the results seem to suggest that it is affirmative, we wish to see more examples to confirm it further.

\begin{acknowledgments}
The authors would like to thank N. Morisawa for his contribution  in the early stage of this work concerning the analysis of the SHEL experiment, 
and also S. Hanashiro, K. Matsuhisa for valuable discussions and comments.  This work was supported in part by the Grant-in-Aid for Scientific Research (KAKENHI), No.~18H03466 and No.~18K13468.
\end{acknowledgments}

\appendix
\section{Position Shifts in the Standard Measurement}
\label{Conventional Measurement}

We first derive the formula \eqref{shiftone} of the pointer shifts in the standard von Neumann measurement.  

Prior to the interaction, the combined system is assumed to be in the product state 
$| \Psi \rangle = | \psi \rangle | \chi \rangle$, where
$| \psi \rangle \in \mathcal{H}$ is the state of the system and $| \chi \rangle \in \mathcal{K}$ is the state of the meter.
The final state of the combined system is
\begin{align}
| \Phi \rangle = e^{-i\frac{\theta}{\hbar}A \otimes Y}| \psi \rangle | \chi \rangle.
\end{align}

The expectation value of the pointer after the interaction then reads
\begin{align}
\langle \Phi | I \otimes X | \Phi \rangle 
= \langle \Psi | e^{i\frac{\theta}{\hbar}A \otimes Y} (I \otimes X) e^{-i\frac{\theta}{\hbar}A \otimes Y} | \Psi \rangle
\end{align}
Using the Baker-Campbell-Hausdorff formula,
\begin{align}
e^P Q e^{-P} = e^{ad_P} Q := Q +  [P, \, Q] + {1\over{2!}}[P, \, [P, \, Q]] + \cdots
\end{align}
we obtain
\begin{align}
&e^{i\frac{\theta}{\hbar}A \otimes Y} (I \otimes X) e^{-i\frac{\theta}{\hbar}A \otimes Y} \nonumber\\
&\qquad = I \otimes X + i\frac{\theta}{\hbar}[A \otimes Y, \, I \otimes X] \nonumber\\
&\qquad = I \otimes X + \theta A \otimes I
\end{align}
on account of $ [ Y, \, X ] = -i\hbar$.  This implies
\begin{align}
\langle \Phi | I \otimes X | \Phi \rangle
&= \langle \Psi | I \otimes X | \Psi \rangle + \theta \langle \Psi | A \otimes I | \Psi \rangle\nonumber\\
&= \langle \Psi | I \otimes X | \Psi \rangle + \theta \langle \psi | A | \psi \rangle
\end{align}
from which we obtain
\begin{align}
\Delta(\theta)
&:= \langle \Phi | I \otimes X | \Phi \rangle - \langle \Psi | I \otimes X | \Psi \rangle \nonumber\\
&=\theta \langle \psi | A | \psi \rangle
\end{align}
which is \eqref{shiftone}.  Note that this result holds for all orders of the coupling $\theta$.

\section{Position Shifts in the Weak Measurement}
\label{Weak Measurement}

We next provide the formula of the position shift in the weak measurement mentioned in \eqref{def:shift3}.

Recall first that $\ket{\xi}$ is the state of the meter after the postselection,
\begin{align}
\label{postselstap}
\ket{\xi} = \frac{\bra{\phi}U(\theta)\ket{\psi} |\chi\rangle}{||\bra{\phi}U(\theta)\ket{\psi}\ket{\chi}||}.
\end{align}
With the possible time development $\ket{\zeta} = V_1\ket{\xi}$ with a unitary operator $V_1$, we obtain
the expectation value of the time developed state,
\begin{align}
\braket{\zeta|X|\zeta}=\bra{\chi}\frac{\braket{\psi|U|\phi}V^{\dag}_1 X V_1\braket{\phi|U|\psi}}{||\braket{\phi|U|\psi}\ket{\chi}||^{2}}\ket{\chi},
\end{align}
where $U$ describes the measurement interaction given by \eqref{def:unitary_interaction}. 

Considering terms up to the linear order in $\theta$, 
\begin{align}
U&=e^{-i\frac{\theta}{\hbar}A\otimes Y}\simeq1-i\frac{\theta}{\hbar} A\otimes Y,
\end{align}
the expectation value becomes
\begin{align}
&\braket{\zeta|X|\zeta}\nonumber\\
&\simeq\bra{\chi}\frac{\braket{\psi|(1+i\frac{\theta}{\hbar}A\otimes Y)|\phi}V^{\dag}_1 X V_1\braket{\phi|(1-i\frac{\theta}{\hbar}A\otimes Y)|\psi}}{||\braket{\phi|(1-i\frac{\theta}{\hbar}A\otimes Y)|\psi}\ket{\chi}||^{2}}\ket{\chi}\nonumber\\
&\simeq\bra{\chi}\frac{(V^{\dag}_1 X V_1+i\frac{\theta}{\hbar}A^{\ast}_{\rm w}YV^{\dag}_1 X V_1-i\frac{\theta}{\hbar}A_{\rm w}V^{\dag}_1 X V_1Y)}{\braket{\chi|(1+i\frac{\theta}{\hbar}A^{\ast}_{\rm w}Y-i\frac{\theta}{\hbar}A_{\rm w}Y)|\chi}}\ket{\chi}\nonumber\\
&\simeq\left(\braket{\chi|V^{\dag}_1 X V_1|\chi}+2\frac{\theta}{\hbar}{\rm Im}[A_{\rm w}\braket{\chi|V^{\dag}_1 X V_1Y|\chi}]\right)\nonumber\\
&\quad\phantom{}\cdot\left(1-2\frac{\theta}{\hbar}{\rm Im}[A_{\rm w}\braket{\chi|Y|\chi}]\right)\nonumber\\
&\simeq\braket{\chi|V^{\dag}_1 X V_1|\chi}\nonumber\\
&\quad\phantom{}+2\frac{\theta}{\hbar}{\rm Im}\left[A_{\rm w}(\braket{\chi|V^{\dag}_1 X V_1Y|\chi}-\braket{\chi|V^{\dag}_1 X V_1|\chi}\braket{\chi|Y|\chi})\right]\nonumber\\
&= \braket{\chi|V^{\dag}_1 X V_1|\chi}+\theta \cdot \frac{2}{\hbar}{\rm Im}[A_{\rm w}{\rm C}_t],
\end{align}
up to the linear order ignoring $O(\theta^2)$, where $A_{\rm w}$ is the weak value in \eqref{def:weakvalue} and 
${\rm C}_t$ is the modified covariance defined in \eqref{def:covariance}.

From this, we learn that the shift of the meter variable $X$ in the weak measurement is given by
\begin{align}
\Delta_{\rm w}(\theta) 
&:= \braket{\zeta|X|\zeta} -  \braket{\chi |X|\chi}  \nonumber \\
&\!\!\!\!\!\!\!\!\! \!\!\!\!\!\!\!\!\!= \braket{\chi|V^{\dag}_1 X V_1|\chi} -  \braket{\chi |X|\chi}  + \theta \cdot \frac{2}{\hbar}{\rm Im} \left[ A_{\rm w}  {\rm C}_t \right] + O(\theta^2),
\end{align}
which is the formula \eqref{def:shift3}.

\section{Limitless Amplification of Weak Value}\label{app:quantum_conditional_expectation}

Let $A$ be a quantum observable on $\mathcal{H}$ with $\mathrm{dim}\,\mathcal{H} \geq 2$.
One then may choose two states $|\psi\rangle, |\psi^{\perp}\rangle \in \mathcal{H}$ fulfilling
\begin{equation}
\langle \psi | \psi^{\perp}\rangle = 0, \quad \langle \psi | A | \psi^{\perp}\rangle \neq 0.
\end{equation}
Given any $z \in \mathbb{C}$, if we choose
\begin{equation}
|\phi(z)\rangle := \frac{|\psi\rangle + z^\ast \cdot |\psi^{\perp}\rangle}{\sqrt{\| \psi \|^{2} + |z|^{2} \| \psi^{\perp} \|^{2}}},
\end{equation}
for our postselected state $|\phi\rangle = |\phi(z)\rangle$, we find that the weak value becomes
\begin{align}
A_{\rm w}(z) = \frac{\langle \phi(z) | A | \psi\rangle}{\langle \phi(z)| \psi\rangle} 
= \langle \psi | A | \psi \rangle + z \cdot \frac{\langle \psi^{\perp} | A | \psi\rangle}{\| \psi \|^{2}}.
\end{align}
Since $z$ can be chosen arbitrarily, the weak value may take an arbitrary value, implying that it can be amplified
as much as we like.

\section{Monotonicity of the Lower-bound Function}
\label{app:monotonicity}

We prove here that the lower-bound function $r(\delta_{\rm st};N)$
given in \eqref{def:lower-bound_probability} is a monotonically increasing 
function 
\footnote{
A function $f$ is said to be monotonically increasing if $a < b$ implies $f(a) \leq f(b)$.  Specifically, $f$ is called strictly monotonically increasing if $a < b$ implies $f(a) < f(b)$.
}
with respect to all of its parameters $\delta_{\rm st}$, $N$ and $q$.  

With the binomial distribution \eqref{def:binom_distr}, 
we first observe that, as a finite weighted average of monotonically increasing functions \eqref{def:Tchebychev_lower-bound}, monotonicity with respect to $\delta_{\rm st}$ is trivial.  To see the monotonicity with respect to $N$, for convenience we extend formally the range of $n$ in
\eqref{def:binom_coeff} from non-negative integers to integers $n \in \mathbb{Z}$ by defining $\binom{N}{n} = 0$ for negative $n$ and thereby
confirm the validity of the formula,
\begin{align}
&\mathrm{Bi}( n; N + 1, q) - \mathrm{Bi} (n; N, q) \nonumber \\
    &\qquad \qquad = q \Big( \mathrm{Bi} (n-1; N, q) - \mathrm{Bi}( n; N, q )\Big),
\end{align}
for $N \in \mathbb{N}$, which can be directly obtained by a simple application of the recursive formula,
\begin{equation}
\binom{N+1}{n} = \binom{N}{n} + \binom{N}{n -1}
\end{equation}
valid for $N > 0$.  This allows us to rewrite
\begin{align}
& r(\delta_{\rm st};N+1,q) - r(\delta_{\rm st};N,q) \nonumber \\
    &\qquad = \sum_{n=1}^{N+1} T(\delta_{\rm st};n) \Big( \mathrm{Bi} ( n; N+1, q ) - \mathrm{Bi}( n; N, q ) \Big) \nonumber \\
    &\qquad = q \sum_{n=1}^{N+1} T(\delta_{\rm st};n) \Big( \mathrm{Bi} ( n-1; N, q ) - \mathrm{Bi}( n; N, q ) \Big) \nonumber \\
    &\qquad = q \sum_{n=1}^{N} \Big( T(\delta_{\rm st};n+1) - T(\delta_{\rm st};n) \Big) \mathrm{Bi}( n; N, q )\nonumber \\
  &\qquad\qquad  +qT(\delta_{\rm st};1)\mathrm{Bi}(0;N,q),
\end{align}
which is always non-negative due to the 
monotonicity of the function \eqref{def:Tchebychev_lower-bound} with respect to $n$.  Finally, as for $q$, first note that the monotonicity trivially holds for $N=0$, since $r(\delta_{\rm st};0,q) = 0$ is a constant function.  For $N > 0$, we may utilize the formula,
\begin{align}
\frac{\partial \mathrm{Bi}( n; N, q)}{\partial q}
    &= N \Big( \mathrm{Bi}( n - 1; N-1, q) \nonumber \\
    &\qquad\qquad\quad - \mathrm{Bi}( n; N-1, q) \Big)
\end{align}
valid for $N > 0$, which can be readily obtained by a direct application of the formula,
\begin{equation}
n  \binom{N}{n} = N \binom{N-1}{n-1}
\end{equation}
valid for $N > 0$.  This allows us to rewrite
\begin{align}
\frac{\partial r(\delta_{\rm st};N,p)}{\partial q}
    &= \sum_{n=1}^{N} T(\delta_{\rm st};n) \,\frac{\partial \mathrm{Bi}( n; N, q )}{\partial q} \nonumber \\
    &= N \sum_{n=1}^{N} T(\delta_{\rm st};n) \Big( \mathrm{Bi}( n - 1; N-1, q) \nonumber \\
        &\qquad\qquad\qquad\qquad\qquad - \mathrm{Bi}( n; N-1, q) \Big) \nonumber \\
    &= N \sum_{n=1}^{N} \Big( T(\delta_{\rm st};n+1) - T(\delta_{\rm st};n) \Big) \nonumber \\
        &\qquad\qquad\qquad\qquad \times \mathrm{Bi}( n; N-1, q ) \nonumber\\
        &\qquad\qquad+NT(\delta_{\rm st};1)\times \mathrm{Bi}( 0; N-1, q ) 
\end{align}
which is always non-negative due to the monotonicity of the function \eqref{def:Tchebychev_lower-bound} with respect to $n$:  this completes the proof of the desired statement.

\section{Paraxial Approximation}
\label{paraxial}
In the text we have used the unitary operators $V_0$ and $V_1$ to describe the effect of beam propagation and its refraction at the lens. 
These operators are constructed with the help of the Fourier optics \cite{JWGoodman} within the so-called paraxial approximation as explained below.
Since the Fourier optics is classical, we first discuss the quantum-classical correspondence of the electromagnetic field. 

\subsection{Quantum-Classical Correspondence}

Let $\ket{\bm{k}}$ be the momentum eigenstate with the eigenvalue $\bm{k} = (k_{x}, k_{y}, k_{z})$ of a single photon.  An arbitrary single
photon state $\ket{\gamma(t)}$, which serves as a meter state considered in our measurement,  may then be written as
\begin{equation}\label{single-photon_state}
\ket{\gamma(t)}=\int d\bm{k}\ \gamma(\bm{k}, t) \hat a^{\dag}_{\bm{k}}\ket{0},
\end{equation}
in the framework of quantum electrodynamics where the eigenstate is realized by the creation operator $\hat a^{\dag}_{\bm{k}}$
acted on the vacuum state $\ket{0}$ as $\ket{\bm{k}} = \hat a^{\dag}_{\bm{k}}\ket{0}$.

Recall that in quantum electrodynamics a field operator such as the scalar potential admits the expansion,
\begin{equation}\label{field-operator}
\hat \phi(\bm{x}, t)=\int d\bm{k} \ \phi(\bm{k})e^{-i\bm{k}\cdot\bm{x} +i\omega t} \hat a_{\bm{k}} + \phi^{\ast}(\bm{k}) e^{i\bm{k}\cdot\bm{x} -i\omega t}\hat a^{\dag}_{\bm{k}},
\end{equation}
where $\omega = c|\bm{k}|$ with $c$ being the speed of light.
Note that the operators corresponding to the electric field, the magnetic field, and the vector potential, also admit an expansion analogous to \eqref{field-operator}.  To each of these, it is customary to associate the classical field as
\begin{equation}\label{classical-field}
\phi(\bm{x}, t) := \braket{0|\hat \phi(\bm{x}, t)|\gamma}=\int d\bm{k}\, \phi(\bm{k}) e^{-i\bm{k}\cdot\bm{x}+i\omega t}\gamma(\bm{k}, t).
\end{equation}
We also note that, for this case, the coefficient function $\phi(\bm{k})$ depends on $\bm{k}$ only through $|\bm{k}|$.  
Moreover, if we fix the overall momentum $|\bm{k}| = k_0$ (as in the case of the experiments analyzed in the paper), we find that $\phi(\bm{k})$ is virtually constant $\phi(\bm{k}) = \phi_0$, which implies
\begin{equation}\label{corresponding-classical-field}
\phi(\bm{x}, t) = \phi_0 \int d\bm{k}\, e^{-i\bm{k}\cdot\bm{x}+i\omega t}\gamma(\bm{k}, t) = \phi_0\, \gamma(\bm{x}, t).
\end{equation}
It follows that, since $\phi(\bm{x}, t)$ in \eqref{classical-field} satisfies the field equation (which is fulfilled by $\hat \phi(\bm{x}, t)$), so does the wave function $\gamma(\bm{x}, t)$.

\subsection{Propagation Effect}

We introduce the coordinate in which the beam propagates along the $z$ direction and the shift obtained by the interaction \eqref{def:unitary_interaction} takes place in the $x$ direction. 
The beam is approximated by the plane wave in the $z$ direction with momentum $p_{z}$ with a tiny spread of the $p_{x}$ component in the $x$ direction,
which are related to the respective wave numbers by $p_{z} = \hbar k_z$ and $p_{x} = \hbar k_x$.
Assuming that the angular frequency $\omega$ of the beam is fixed,  the profile of the electromagnetic field can be written as
\begin{align}
\phi(\bm{x},t)=g(\bm{x})e^{i(k_0z-\omega t)}.
\label{eq:paraxialapprox}
\end{align}
In this form, the wave equation reads
\begin{align}
0&=\left(\frac{1}{c^{2}}\frac{\partial^{2}}{\partial t^{2}}-\nabla^{2}\right)\phi(\bm{x},t) \label{waveequationofthebeam}\\
&=e^{i(k_0z-\omega t)}\left(-\nabla^{2}-2ik_0\frac{\partial}{\partial z}\right) g(\bm{x}).\label{waveequationofg}
\end{align}

The paraxial approximation is valid when the conditions
\begin{equation}\label{paraxial_approximation}
\left|\frac{\partial^{2}g(\bm{x})}{\partial z^{2}}\right|\ll\left|\frac{\partial^{2}g(\bm{x})}{\partial x^{2}}\right|, \,
\left|\frac{\partial^{2}g(\bm{x})}{\partial y^{2}}\right|, \,
\left| k_0 \frac{\partial g(\bm{x})}{\partial z}\right|,
\end{equation}
are fulfilled.  
To check the validity of these, 
we consider the Fourier expansion of $\phi(\bm{x},t)$,
\begin{equation}\label{f_fourier_transformation}
\phi(\bm{x},t)=\int d^{3}k\, \tilde{\phi}(\bm{k})e^{i(\bm{k}\cdot\bm{x}-\omega t)}.
\end{equation}
Plugging \eqref{f_fourier_transformation} into \eqref{waveequationofthebeam}, we obtain
\begin{equation}\label{condition_of_momentum}
k_x^2+k_y^2+k_z^2=k_0^2,
\end{equation}
for each mode $\bm{k}$. 
Combining \eqref{eq:paraxialapprox} with \eqref{f_fourier_transformation}, one finds
\begin{equation}
\label{paraxial_fourier_rep}
g(\bm{x})=\int d^{3}k\, \tilde{\phi}(\bm{k})e^{i\left(k_x x+k_y y+ (k_z-k_0)z\right)}.
\end{equation}
Then the conditions \eqref{paraxial_approximation} are satisfied if
\begin{align}\label{fourier_paraxial_approx}
|k_z-k_0|\ll |k_x|, \, |k_y|,\, \sqrt{|k_0(k_z-k_0)|},
\end{align}
which are assured if we just have
\begin{align}\label{paraxial_condition}
k_x, k_y\ll k_0
\end{align}
on account of \eqref{condition_of_momentum}. 
The beam used in the experiments which we analyzed in the paper indeed fulfills these conditions \eqref{paraxial_condition}.

Within the paraxial approximation \eqref{paraxial_approximation}, we therefore obtain from the wave equation \eqref{waveequationofg},
\begin{equation}\label{paraxial_wave_equation}
\left(\frac{\partial^{2}}{\partial x^{2}}+\frac{\partial^{2}}{\partial y^{2}}+2ik_0\frac{\partial}{\partial z}\right)g(\bm{x})=0.
\end{equation}
If we use \eqref{paraxial_fourier_rep}, we find that  the equation \eqref{paraxial_wave_equation} can be solved as
\begin{align}
g(\bm{x})=\int d^3k\, \tilde{\phi}(\bm{k})e^{i\left(k_x x+k_y y-\frac{1}{2k_0}(k_x^2+k_y^2)z\right)},
\end{align}
which implies
\begin{align}
\label{beamprofile}
\phi(\bm{x},t)=\int d^3k\, \tilde{\phi}(\bm{k})e^{i\left(k_x x+k_y y-\frac{1}{2k_0}(k_x^2+k_y^2)z\right)}e^{i(k_0z-\omega t)}.
\end{align}
Now, in view of \eqref{corresponding-classical-field}, one may introduce the unitary operator representing the propagation effect,
\begin{equation}
V_{\rm P} =e^{-i\frac{\hat{k}^{2}_{x} + \hat{k}^{2}_{y}}{2k_0}l},
\end{equation}
and thereby express the relation between 
$\phi(x,y,0, t)$ to $\phi(x,y,l, t)$ in terms of the wave function $\gamma(\bm{x}, t)$ as 
\begin{equation}
\label{propeffect}
\gamma(x,y,l, t) = \braket{x,y,l|\gamma} =\bra{x,y,0} V_{\rm P}\ket{\gamma}.
\end{equation}
The unitary operator (\ref{vone},\ref{vone2}) used in the text arises when we have the shift only in the $x$ direction.  The state
appearing above corresponds to the meter state $\ket{\xi}$ there.

\subsection{Lens Effect}

In addition to the propagation effect, we also need to take into account the effect of lens used in the experiment.
Since the the profile function \eqref{beamprofile} of the laser beam is a superposition of various plane waves 
with fixed $\bm{k}$, it is enough for us to consider the effect only on the plane wave.  

\begin{figure}[h]
\label{lens}
\includegraphics[width=8cm]{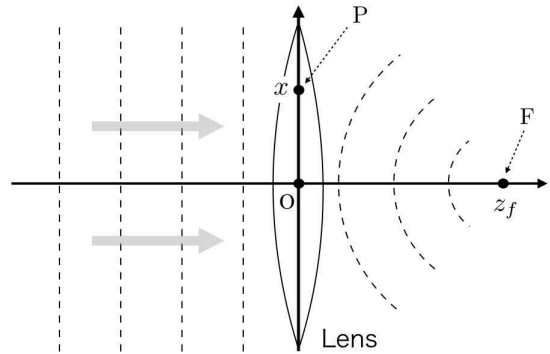}
\caption{Lens effect in the paraxial approximation.}
\end{figure}

For this, we first observe that, during the passage of the plane wave from the point P in the lens to the focal point F (see Fig.\ref{lens}),
it acquires the extra phase,
\begin{equation}\label{phase-dif}
\vartheta(x)=k_0 \left(\sqrt{z_{f}^{2}+x^{2}}-z_{f}\right),
\end{equation}
compared to the passage in the center from $O$ to $F$,
where $z_{f}$ is the focal length of the lens which are assumed to be thin enough so that the thickness effect can be ignored
in our calculation of the extra phase.  For $z_{f}\gg x$, we then have
\begin{equation}
\vartheta(x) \simeq k_0 \frac{x^{2}}{2z_{f}},
\end{equation}
which yields the phase change $e^{-ik_0\frac{x^{2}}{2z_{f}}}$ for each of the plane waves.  
As a result, the effect of the lens can be incorporated if we modify the profile function \eqref{beamprofile},
or equivalently the wave function $\gamma(\bm{x}, t)$, by inserting the unitary operator representing the lens effect,
\begin{equation}
V_{\rm L}=e^{-i\frac{k_0}{2z_{f}}\hat{x}^{2}},
\end{equation}
at an appropriate position as we have done in the formula \eqref{propeffect} in the case of the propagation effect.

\bibliography{main}

\begin{thebibliography}{28}%
\makeatletter
\providecommand \@ifxundefined [1]{%
 \@ifx{#1\undefined}
}%
\providecommand \@ifnum [1]{%
 \ifnum #1\expandafter \@firstoftwo
 \else \expandafter \@secondoftwo
 \fi
}%
\providecommand \@ifx [1]{%
 \ifx #1\expandafter \@firstoftwo
 \else \expandafter \@secondoftwo
 \fi
}%
\providecommand \natexlab [1]{#1}%
\providecommand \enquote  [1]{``#1''}%
\providecommand \bibnamefont  [1]{#1}%
\providecommand \bibfnamefont [1]{#1}%
\providecommand \citenamefont [1]{#1}%
\providecommand \href@noop [0]{\@secondoftwo}%
\providecommand \href [0]{\begingroup \@sanitize@url \@href}%
\providecommand \@href[1]{\@@startlink{#1}\@@href}%
\providecommand \@@href[1]{\endgroup#1\@@endlink}%
\providecommand \@sanitize@url [0]{\catcode `\\12\catcode `\$12\catcode
  `\&12\catcode `\#12\catcode `\^12\catcode `\_12\catcode `\%12\relax}%
\providecommand \@@startlink[1]{}%
\providecommand \@@endlink[0]{}%
\providecommand \url  [0]{\begingroup\@sanitize@url \@url }%
\providecommand \@url [1]{\endgroup\@href {#1}{\urlprefix }}%
\providecommand \urlprefix  [0]{URL }%
\providecommand \Eprint [0]{\href }%
\providecommand \doibase [0]{http://dx.doi.org/}%
\providecommand \selectlanguage [0]{\@gobble}%
\providecommand \bibinfo  [0]{\@secondoftwo}%
\providecommand \bibfield  [0]{\@secondoftwo}%
\providecommand \translation [1]{[#1]}%
\providecommand \BibitemOpen [0]{}%
\providecommand \bibitemStop [0]{}%
\providecommand \bibitemNoStop [0]{.\EOS\space}%
\providecommand \EOS [0]{\spacefactor3000\relax}%
\providecommand \BibitemShut  [1]{\csname bibitem#1\endcsname}%
\let\auto@bib@innerbib\@empty
\bibitem [{\citenamefont {Hosten}\ and\ \citenamefont
  {Kwiat}(2008)}]{Hosten_2008}%
  \BibitemOpen
  \bibfield  {author} {\bibinfo {author} {\bibfnamefont {O.}~\bibnamefont
  {Hosten}}\ and\ \bibinfo {author} {\bibfnamefont {P.}~\bibnamefont {Kwiat}},\
  }\href {\doibase 10.1126/science.1152697} {\bibfield  {journal} {\bibinfo
  {journal} {Science}\ }\textbf {\bibinfo {volume} {319}},\ \bibinfo {pages}
  {787} (\bibinfo {year} {2008})}\BibitemShut {NoStop}%
\bibitem [{\citenamefont {Dixon}\ \emph {et~al.}(2009)\citenamefont {Dixon},
  \citenamefont {Starling}, \citenamefont {Jordan},\ and\ \citenamefont
  {Howell}}]{Dixon_2009}%
  \BibitemOpen
  \bibfield  {author} {\bibinfo {author} {\bibfnamefont {P.~B.}\ \bibnamefont
  {Dixon}}, \bibinfo {author} {\bibfnamefont {D.~J.}\ \bibnamefont {Starling}},
  \bibinfo {author} {\bibfnamefont {A.~N.}\ \bibnamefont {Jordan}}, \ and\
  \bibinfo {author} {\bibfnamefont {J.~C.}\ \bibnamefont {Howell}},\ }\href
  {\doibase 10.1103/PhysRevLett.102.173601} {\bibfield  {journal} {\bibinfo
  {journal} {Phys. Rev. Lett.}\ }\textbf {\bibinfo {volume} {102}},\ \bibinfo
  {pages} {173601} (\bibinfo {year} {2009})}\BibitemShut {NoStop}%
\bibitem [{\citenamefont {Aharonov}\ \emph {et~al.}(1964)\citenamefont
  {Aharonov}, \citenamefont {Bergmann},\ and\ \citenamefont
  {Lebowitz}}]{Aharonov_1964}%
  \BibitemOpen
  \bibfield  {author} {\bibinfo {author} {\bibfnamefont {Y.}~\bibnamefont
  {Aharonov}}, \bibinfo {author} {\bibfnamefont {P.~G.}\ \bibnamefont
  {Bergmann}}, \ and\ \bibinfo {author} {\bibfnamefont {L.}~\bibnamefont
  {Lebowitz}},\ }\href@noop {} {\bibfield  {journal} {\bibinfo  {journal}
  {Phys. Rev.}\ }\textbf {\bibinfo {volume} {134}},\ \bibinfo {pages} {B1410}
  (\bibinfo {year} {1964})}\BibitemShut {NoStop}%
\bibitem [{\citenamefont {Aharonov}\ \emph {et~al.}(1988)\citenamefont
  {Aharonov}, \citenamefont {Albert},\ and\ \citenamefont
  {Vaidman}}]{Aharonov_1988}%
  \BibitemOpen
  \bibfield  {author} {\bibinfo {author} {\bibfnamefont {Y.}~\bibnamefont
  {Aharonov}}, \bibinfo {author} {\bibfnamefont {D.~Z.}\ \bibnamefont
  {Albert}}, \ and\ \bibinfo {author} {\bibfnamefont {L.}~\bibnamefont
  {Vaidman}},\ }\href@noop {} {\bibfield  {journal} {\bibinfo  {journal} {Phys.
  Rev. Lett.}\ }\textbf {\bibinfo {volume} {60}},\ \bibinfo {pages} {1351}
  (\bibinfo {year} {1988})}\BibitemShut {NoStop}%
\bibitem [{\citenamefont {Kocsis}\ \emph {et~al.}(2011)\citenamefont {Kocsis},
  \citenamefont {Braverman}, \citenamefont {Ravets}, \citenamefont {Stevens},
  \citenamefont {Mirin}, \citenamefont {Shalm},\ and\ \citenamefont
  {Steinberg}}]{Kocsis1170}%
  \BibitemOpen
  \bibfield  {author} {\bibinfo {author} {\bibfnamefont {S.}~\bibnamefont
  {Kocsis}}, \bibinfo {author} {\bibfnamefont {B.}~\bibnamefont {Braverman}},
  \bibinfo {author} {\bibfnamefont {S.}~\bibnamefont {Ravets}}, \bibinfo
  {author} {\bibfnamefont {M.~J.}\ \bibnamefont {Stevens}}, \bibinfo {author}
  {\bibfnamefont {R.~P.}\ \bibnamefont {Mirin}}, \bibinfo {author}
  {\bibfnamefont {L.~K.}\ \bibnamefont {Shalm}}, \ and\ \bibinfo {author}
  {\bibfnamefont {A.~M.}\ \bibnamefont {Steinberg}},\ }\href {\doibase
  10.1126/science.1202218} {\bibfield  {journal} {\bibinfo  {journal}
  {Science}\ }\textbf {\bibinfo {volume} {332}},\ \bibinfo {pages} {1170}
  (\bibinfo {year} {2011})}\BibitemShut {NoStop}%
\bibitem [{\citenamefont {Mori}\ and\ \citenamefont
  {Tsutsui}(2015)}]{Mori2015}%
  \BibitemOpen
  \bibfield  {author} {\bibinfo {author} {\bibfnamefont {T.}~\bibnamefont
  {Mori}}\ and\ \bibinfo {author} {\bibfnamefont {I.}~\bibnamefont {Tsutsui}},\
  }\href {\doibase 10.1007/s40509-015-0039-5} {\bibfield  {journal} {\bibinfo
  {journal} {Quantum Stud.: Math. Found.}\ }\textbf {\bibinfo {volume} {2}},\
  \bibinfo {pages} {371} (\bibinfo {year} {2015})}\BibitemShut {NoStop}%
\bibitem [{\citenamefont {Aharonov}\ and\ \citenamefont
  {Vaidman}(1991)}]{Aharonov_1991}%
  \BibitemOpen
  \bibfield  {author} {\bibinfo {author} {\bibfnamefont {Y.}~\bibnamefont
  {Aharonov}}\ and\ \bibinfo {author} {\bibfnamefont {L.}~\bibnamefont
  {Vaidman}},\ }\href@noop {} {\bibfield  {journal} {\bibinfo  {journal} {J.
  Phys. A: Math. Gen.}\ }\textbf {\bibinfo {volume} {24}},\ \bibinfo {pages}
  {2315} (\bibinfo {year} {1991})}\BibitemShut {NoStop}%
\bibitem [{\citenamefont {Aharonov}\ \emph {et~al.}(2002)\citenamefont
  {Aharonov}, \citenamefont {Botero}, \citenamefont {Popescu}, \citenamefont
  {Reznik},\ and\ \citenamefont {Tollaksen}}]{Aharonov_2002}%
  \BibitemOpen
  \bibfield  {author} {\bibinfo {author} {\bibfnamefont {Y.}~\bibnamefont
  {Aharonov}}, \bibinfo {author} {\bibfnamefont {A.}~\bibnamefont {Botero}},
  \bibinfo {author} {\bibfnamefont {S.}~\bibnamefont {Popescu}}, \bibinfo
  {author} {\bibfnamefont {B.}~\bibnamefont {Reznik}}, \ and\ \bibinfo {author}
  {\bibfnamefont {J.}~\bibnamefont {Tollaksen}},\ }\href {\doibase
  10.1016/S0375-9601(02)00986-6} {\bibfield  {journal} {\bibinfo  {journal}
  {Phys. Lett. A}\ }\textbf {\bibinfo {volume} {301}},\ \bibinfo {pages} {130 }
  (\bibinfo {year} {2002})}\BibitemShut {NoStop}%
\bibitem [{\citenamefont {Aharonov}\ \emph {et~al.}(2013)\citenamefont
  {Aharonov}, \citenamefont {Popescu}, \citenamefont {Rohrlich},\ and\
  \citenamefont {Skrzypczyk}}]{Aharonov_2013}%
  \BibitemOpen
  \bibfield  {author} {\bibinfo {author} {\bibfnamefont {Y.}~\bibnamefont
  {Aharonov}}, \bibinfo {author} {\bibfnamefont {S.}~\bibnamefont {Popescu}},
  \bibinfo {author} {\bibfnamefont {D.}~\bibnamefont {Rohrlich}}, \ and\
  \bibinfo {author} {\bibfnamefont {P.}~\bibnamefont {Skrzypczyk}},\ }\href
  {\doibase 10.1088/1367-2630/15/11/113015} {\bibfield  {journal} {\bibinfo
  {journal} {New J. Phys.}\ }\textbf {\bibinfo {volume} {15}},\ \bibinfo
  {pages} {113015} (\bibinfo {year} {2013})}\BibitemShut {NoStop}%
\bibitem [{\citenamefont {Yokota}\ \emph {et~al.}(2009)\citenamefont {Yokota},
  \citenamefont {Yamamoto}, \citenamefont {Koashi},\ and\ \citenamefont
  {Imoto}}]{Yokota_2009}%
  \BibitemOpen
  \bibfield  {author} {\bibinfo {author} {\bibfnamefont {K.}~\bibnamefont
  {Yokota}}, \bibinfo {author} {\bibfnamefont {T.}~\bibnamefont {Yamamoto}},
  \bibinfo {author} {\bibfnamefont {M.}~\bibnamefont {Koashi}}, \ and\ \bibinfo
  {author} {\bibfnamefont {N.}~\bibnamefont {Imoto}},\ }\href
  {http://stacks.iop.org/1367-2630/11/i=3/a=033011} {\bibfield  {journal}
  {\bibinfo  {journal} {New J. Phys.}\ }\textbf {\bibinfo {volume} {11}},\
  \bibinfo {pages} {033011} (\bibinfo {year} {2009})}\BibitemShut {NoStop}%
\bibitem [{\citenamefont {Denkmayr}\ \emph {et~al.}(2014)\citenamefont
  {Denkmayr}, \citenamefont {Geppert}, \citenamefont {Sponar}, \citenamefont
  {Lemmel}, \citenamefont {Matzkin}, \citenamefont {Tollaksen},\ and\
  \citenamefont {Hasegawa}}]{Denkmayr:2014aa}%
  \BibitemOpen
  \bibfield  {author} {\bibinfo {author} {\bibfnamefont {T.}~\bibnamefont
  {Denkmayr}}, \bibinfo {author} {\bibfnamefont {H.}~\bibnamefont {Geppert}},
  \bibinfo {author} {\bibfnamefont {S.}~\bibnamefont {Sponar}}, \bibinfo
  {author} {\bibfnamefont {H.}~\bibnamefont {Lemmel}}, \bibinfo {author}
  {\bibfnamefont {A.}~\bibnamefont {Matzkin}}, \bibinfo {author} {\bibfnamefont
  {J.}~\bibnamefont {Tollaksen}}, \ and\ \bibinfo {author} {\bibfnamefont
  {Y.}~\bibnamefont {Hasegawa}},\ }\href {https://doi.org/10.1038/ncomms5492}
  {\bibfield  {journal} {\bibinfo  {journal} {Nat. Commun.}\ }\textbf {\bibinfo
  {volume} {5}},\ \bibinfo {pages} {4492 EP } (\bibinfo {year}
  {2014})}\BibitemShut {NoStop}%
\bibitem [{\citenamefont {Okamoto}\ and\ \citenamefont
  {Takeuchi}(2016)}]{Okamoto:2016aa}%
  \BibitemOpen
  \bibfield  {author} {\bibinfo {author} {\bibfnamefont {R.}~\bibnamefont
  {Okamoto}}\ and\ \bibinfo {author} {\bibfnamefont {S.}~\bibnamefont
  {Takeuchi}},\ }\href {https://doi.org/10.1038/srep35161} {\bibfield
  {journal} {\bibinfo  {journal} {Sci. Rep.}\ }\textbf {\bibinfo {volume}
  {6}},\ \bibinfo {pages} {35161 EP } (\bibinfo {year} {2016})}\BibitemShut
  {NoStop}%
\bibitem [{\citenamefont {Danan}\ \emph {et~al.}(2013)\citenamefont {Danan},
  \citenamefont {Farfurnik}, \citenamefont {Bar-Ad},\ and\ \citenamefont
  {Vaidman}}]{Danan_2013}%
  \BibitemOpen
  \bibfield  {author} {\bibinfo {author} {\bibfnamefont {A.}~\bibnamefont
  {Danan}}, \bibinfo {author} {\bibfnamefont {D.}~\bibnamefont {Farfurnik}},
  \bibinfo {author} {\bibfnamefont {S.}~\bibnamefont {Bar-Ad}}, \ and\ \bibinfo
  {author} {\bibfnamefont {L.}~\bibnamefont {Vaidman}},\ }\href {\doibase
  10.1103/PhysRevLett.111.240402} {\bibfield  {journal} {\bibinfo  {journal}
  {Phys. Rev. Lett.}\ }\textbf {\bibinfo {volume} {111}},\ \bibinfo {pages}
  {240402} (\bibinfo {year} {2013})}\BibitemShut {NoStop}%
\bibitem [{\citenamefont {Vaidman}\ and\ \citenamefont
  {Tsutsui}(2018)}]{Vaidman_2018}%
  \BibitemOpen
  \bibfield  {author} {\bibinfo {author} {\bibfnamefont {L.}~\bibnamefont
  {Vaidman}}\ and\ \bibinfo {author} {\bibfnamefont {I.}~\bibnamefont
  {Tsutsui}},\ }\href {http://www.mdpi.com/1099-4300/20/7/538} {\bibfield
  {journal} {\bibinfo  {journal} {Entropy}\ }\textbf {\bibinfo {volume} {20}},\
  \bibinfo {pages} {538} (\bibinfo {year} {2018})}\BibitemShut {NoStop}%
\bibitem [{\citenamefont {Sinclair}\ \emph {et~al.}(2018)\citenamefont
  {Sinclair}, \citenamefont {Spierings}, \citenamefont {Brodutch},\ and\
  \citenamefont {Steinberg}}]{1808.09951}%
  \BibitemOpen
  \bibfield  {author} {\bibinfo {author} {\bibfnamefont {J.}~\bibnamefont
  {Sinclair}}, \bibinfo {author} {\bibfnamefont {D.}~\bibnamefont {Spierings}},
  \bibinfo {author} {\bibfnamefont {A.}~\bibnamefont {Brodutch}}, \ and\
  \bibinfo {author} {\bibfnamefont {A.~M.}\ \bibnamefont {Steinberg}},\
  }\href@noop {} {\  (\bibinfo {year} {2018})},\ \Eprint
  {http://arxiv.org/abs/arXiv:1808.09951} {arXiv:1808.09951} \BibitemShut
  {NoStop}%
\bibitem [{\citenamefont {Georgiev}\ and\ \citenamefont
  {Cohen}(2018)}]{1810.05039}%
  \BibitemOpen
  \bibfield  {author} {\bibinfo {author} {\bibfnamefont {D.}~\bibnamefont
  {Georgiev}}\ and\ \bibinfo {author} {\bibfnamefont {E.}~\bibnamefont
  {Cohen}},\ }\href@noop {} {\  (\bibinfo {year} {2018})},\ \Eprint
  {http://arxiv.org/abs/arXiv:1810.05039} {arXiv:1810.05039} \BibitemShut
  {NoStop}%
\bibitem [{\citenamefont {Thekkadath}\ \emph {et~al.}(2018)\citenamefont
  {Thekkadath}, \citenamefont {Hufnagel},\ and\ \citenamefont
  {Lundeen}}]{Thekkadath_2018}%
  \BibitemOpen
  \bibfield  {author} {\bibinfo {author} {\bibfnamefont {G.~S.}\ \bibnamefont
  {Thekkadath}}, \bibinfo {author} {\bibfnamefont {F.}~\bibnamefont
  {Hufnagel}}, \ and\ \bibinfo {author} {\bibfnamefont {J.~S.}\ \bibnamefont
  {Lundeen}},\ }\href {http://stacks.iop.org/1367-2630/20/i=11/a=113034}
  {\bibfield  {journal} {\bibinfo  {journal} {New J. Phys.}\ }\textbf {\bibinfo
  {volume} {20}},\ \bibinfo {pages} {113034} (\bibinfo {year}
  {2018})}\BibitemShut {NoStop}%
\bibitem [{\citenamefont {Dressel}\ \emph {et~al.}(2014)\citenamefont
  {Dressel}, \citenamefont {Malik}, \citenamefont {Miatto}, \citenamefont
  {Jordan},\ and\ \citenamefont {Boyd}}]{Dressel_2014}%
  \BibitemOpen
  \bibfield  {author} {\bibinfo {author} {\bibfnamefont {J.}~\bibnamefont
  {Dressel}}, \bibinfo {author} {\bibfnamefont {M.}~\bibnamefont {Malik}},
  \bibinfo {author} {\bibfnamefont {F.~M.}\ \bibnamefont {Miatto}}, \bibinfo
  {author} {\bibfnamefont {A.~N.}\ \bibnamefont {Jordan}}, \ and\ \bibinfo
  {author} {\bibfnamefont {R.~W.}\ \bibnamefont {Boyd}},\ }\href {\doibase
  10.1103/RevModPhys.86.307} {\bibfield  {journal} {\bibinfo  {journal} {Rev.
  Mod. Phys.}\ }\textbf {\bibinfo {volume} {86}},\ \bibinfo {pages} {307}
  (\bibinfo {year} {2014})}\BibitemShut {NoStop}%
\bibitem [{\citenamefont {{Chun-Wang Wu}}\ \emph {et~al.}(2018)\citenamefont
  {{Chun-Wang Wu}}, \citenamefont {{Jie Zhang}}, \citenamefont {{Yi Xie}},
  \citenamefont {{Bao-Quan Ou}}, \citenamefont {{Wei Wu}},\ and\ \citenamefont
  {{Ping-Xing Chen}}}]{1811.06170}%
  \BibitemOpen
  \bibfield  {author} {\bibinfo {author} {\bibnamefont {{Chun-Wang Wu}}},
  \bibinfo {author} {\bibnamefont {{Jie Zhang}}}, \bibinfo {author}
  {\bibnamefont {{Yi Xie}}}, \bibinfo {author} {\bibnamefont {{Bao-Quan Ou}}},
  \bibinfo {author} {\bibnamefont {{Wei Wu}}}, \ and\ \bibinfo {author}
  {\bibnamefont {{Ping-Xing Chen}}},\ }\href@noop {} {\  (\bibinfo {year}
  {2018})},\ \Eprint {http://arxiv.org/abs/arXiv:1811.06170} {arXiv:1811.06170}
  \BibitemShut {NoStop}%
\bibitem [{\citenamefont {Nakamura}\ and\ \citenamefont
  {Fujimoto}(2018)}]{NAKAMURA_2018}%
  \BibitemOpen
  \bibfield  {author} {\bibinfo {author} {\bibfnamefont {K.}~\bibnamefont
  {Nakamura}}\ and\ \bibinfo {author} {\bibfnamefont {M.}~\bibnamefont
  {Fujimoto}},\ }\href {\doibase https://doi.org/10.1016/j.aop.2018.03.009}
  {\bibfield  {journal} {\bibinfo  {journal} {Annals of Physics}\ }\textbf
  {\bibinfo {volume} {392}},\ \bibinfo {pages} {71 } (\bibinfo {year}
  {2018})}\BibitemShut {NoStop}%
\bibitem [{\citenamefont {Kawana}\ and\ \citenamefont
  {Ueda}(2018)}]{Kawana_2018}%
  \BibitemOpen
  \bibfield  {author} {\bibinfo {author} {\bibfnamefont {K.}~\bibnamefont
  {Kawana}}\ and\ \bibinfo {author} {\bibfnamefont {D.}~\bibnamefont {Ueda}},\
  }\href@noop {} {\  (\bibinfo {year} {2018})},\ \Eprint
  {http://arxiv.org/abs/arXiv:1804.09505} {arXiv:1804.09505} \BibitemShut
  {NoStop}%
\bibitem [{\citenamefont {Knee}\ and\ \citenamefont
  {Gauger}(2014)}]{Knee_2014}%
  \BibitemOpen
  \bibfield  {author} {\bibinfo {author} {\bibfnamefont {G.~C.}\ \bibnamefont
  {Knee}}\ and\ \bibinfo {author} {\bibfnamefont {E.~M.}\ \bibnamefont
  {Gauger}},\ }\href {\doibase 10.1103/PhysRevX.4.011032} {\bibfield  {journal}
  {\bibinfo  {journal} {Phys. Rev. X}\ }\textbf {\bibinfo {volume} {4}},\
  \bibinfo {pages} {011032} (\bibinfo {year} {2014})}\BibitemShut {NoStop}%
\bibitem [{\citenamefont {Koike}\ and\ \citenamefont
  {Tanaka}(2011)}]{koiketanaka_11}%
  \BibitemOpen
  \bibfield  {author} {\bibinfo {author} {\bibfnamefont {T.}~\bibnamefont
  {Koike}}\ and\ \bibinfo {author} {\bibfnamefont {S.}~\bibnamefont {Tanaka}},\
  }\href {\doibase 10.1103/PhysRevA.84.062106} {\bibfield  {journal} {\bibinfo
  {journal} {Phys. Rev. A}\ }\textbf {\bibinfo {volume} {84}},\ \bibinfo
  {pages} {062106} (\bibinfo {year} {2011})}\BibitemShut {NoStop}%
\bibitem [{\citenamefont {Lee}\ and\ \citenamefont {Tsutsui}(2014)}]{Lee_2014}%
  \BibitemOpen
  \bibfield  {author} {\bibinfo {author} {\bibfnamefont {J.}~\bibnamefont
  {Lee}}\ and\ \bibinfo {author} {\bibfnamefont {I.}~\bibnamefont {Tsutsui}},\
  }\href {\doibase 10.1007/s40509-014-0002-x} {\bibfield  {journal} {\bibinfo
  {journal} {Quantum Stud.: Math. Found.}\ }\textbf {\bibinfo {volume} {1}},\
  \bibinfo {pages} {65} (\bibinfo {year} {2014})}\BibitemShut {NoStop}%
\bibitem [{\citenamefont {Lee}\ and\ \citenamefont
  {Tsutsui}(2017)}]{Lee_2016_PTEP}%
  \BibitemOpen
  \bibfield  {author} {\bibinfo {author} {\bibfnamefont {J.}~\bibnamefont
  {Lee}}\ and\ \bibinfo {author} {\bibfnamefont {I.}~\bibnamefont {Tsutsui}},\
  }\href {\doibase 10.1093/ptep/ptx024} {\bibfield  {journal} {\bibinfo
  {journal} {Prog. Theor. Exp. Phys.}\ }\textbf {\bibinfo {volume} {2017}},\
  \bibinfo {pages} {052A01} (\bibinfo {year} {2017})}\BibitemShut {NoStop}%
\bibitem [{\citenamefont {{Young-Wook Cho}}\ \emph {et~al.}(2010)\citenamefont
  {{Young-Wook Cho}}, \citenamefont {{Hyang-Tag Lim}}, \citenamefont
  {{Young-Sik Ra}},\ and\ \citenamefont {{Yoon-Ho Kim}}}]{Cho_2010}%
  \BibitemOpen
  \bibfield  {author} {\bibinfo {author} {\bibnamefont {{Young-Wook Cho}}},
  \bibinfo {author} {\bibnamefont {{Hyang-Tag Lim}}}, \bibinfo {author}
  {\bibnamefont {{Young-Sik Ra}}}, \ and\ \bibinfo {author} {\bibnamefont
  {{Yoon-Ho Kim}}},\ }\href {http://stacks.iop.org/1367-2630/12/i=2/a=023036}
  {\bibfield  {journal} {\bibinfo  {journal} {New J. Phys.}\ }\textbf {\bibinfo
  {volume} {12}},\ \bibinfo {pages} {023036} (\bibinfo {year}
  {2010})}\BibitemShut {NoStop}%
\bibitem [{\citenamefont {Jordan}\ \emph {et~al.}(2014)\citenamefont {Jordan},
  \citenamefont {{J. Martinez-Rincon}},\ and\ \citenamefont
  {Howell}}]{Jordan_2014}%
  \BibitemOpen
  \bibfield  {author} {\bibinfo {author} {\bibfnamefont {A.~N.}\ \bibnamefont
  {Jordan}}, \bibinfo {author} {\bibnamefont {{J. Martinez-Rincon}}}, \ and\
  \bibinfo {author} {\bibfnamefont {J.~C.}\ \bibnamefont {Howell}},\ }\href
  {\doibase 10.1103/PhysRevX.4.011031} {\bibfield  {journal} {\bibinfo
  {journal} {Phys. Rev. X}\ }\textbf {\bibinfo {volume} {4}},\ \bibinfo {pages}
  {011031} (\bibinfo {year} {2014})}\BibitemShut {NoStop}%
\bibitem [{\citenamefont {Goodman}(1996)}]{JWGoodman}%
  \BibitemOpen
  \bibfield  {author} {\bibinfo {author} {\bibfnamefont {J.~W.}\ \bibnamefont
  {Goodman}},\ }\href@noop {} {\emph {\bibinfo {title} {Introduction to Fourier
  Optics}}},\ \bibinfo {edition} {2nd}\ ed.\ (\bibinfo  {publisher}
  {McGraw-Hill},\ \bibinfo {year} {1996})\BibitemShut {NoStop}%
\end{thebibliography}%

\end{document}